\documentclass[%
superscriptaddress,
preprint,
preprintnumbers,
nofootinbib,
 amsmath,amssymb,
 aps,
prd,
dvipsnames,
floatfix,
]{revtex4-1}

\usepackage{setspace}
\usepackage{bm}
\usepackage{pstricks}
\usepackage{lineno}
\usepackage{makecell}
\usepackage{verbatim}
   
  \usepackage{calc}               
  \usepackage{fancybox}           
  \usepackage{fancyhdr}           
  \usepackage{fancyvrb}           
  \usepackage{graphicx}           
  \usepackage[colorlinks=true,citecolor=blue,linkcolor=blue,urlcolor=blue]{hyperref}
  \usepackage{multido}            
  \usepackage{pifont}             
  \usepackage{pst-3d}             
  \usepackage{pst-grad}           
  \usepackage{pst-node}           
  \usepackage{url}                
  \usepackage{dcolumn}            
  \usepackage[normalem]{ulem}     
  \def\GeV2{(GeV/c)$^2$}

\newcommand{\CJ}{{\textsc{\textrm{CJ}}}}

\newcommand{\bonus}{BONuS }
\newcommand*\dd{\mathop{}\!\mathrm{d}}

\begin{document}

\preprint{JLAB-THY-23-3907}

\title{Extraction of the neutron $F_{2}$ structure function from inclusive proton and deuteron deep-inelastic scattering data}

\author{S.~Li}
\affiliation{ 
University of New Hampshire, Durham, New Hampshire 03824, USA}
\affiliation{
Lawrence Berkeley National Laboratory, Berkeley, California 94720, USA}
\author{A.~Accardi}
\affiliation{Hampton University, Hampton, Virginia 23668, USA}
\affiliation{Jefferson Lab, Newport News, Virginia 23606, USA}
\author{M.~Cerutti}
\affiliation{Hampton University, Hampton, Virginia 23668, USA}
\affiliation{Jefferson Lab, Newport News, Virginia 23606, USA}
\author{I.~P.~Fernando}
\affiliation{University of Virginia, Charlottesville, Virginia 22904, USA}
\author{C.~E.~Keppel}
\affiliation{Jefferson Lab, Newport News, Virginia 23606, USA}
\author{W.~Melnitchouk}
\affiliation{Jefferson Lab, Newport News, Virginia 23606, USA}
\author{P.~Monaghan}
\affiliation{Christopher Newport University, Newport News, Virginia, 23606, USA}
\author{G.~Niculescu}
\affiliation{James Madison University, Harrisonburg, Virginia 22801, USA}
\author{M.~I.~Niculescu}
\affiliation{James Madison University, Harrisonburg, Virginia 22801, USA}
\author{J.~F.~Owens}
\affiliation{Florida State University, Tallahassee, Florida 32306, USA}


\begin{abstract}
The available world deep-inelastic scattering data on proton and deuteron structure functions $F_2^p$, $F_2^d$, and their ratios, are leveraged to extract the free neutron $F_2^n$ structure function, the $F_2^n/F_2^p$ ratio, and associated uncertainties using the latest nuclear effect calculations in the deuteron.
Special attention is devoted to the normalization of the proton and deuteron experimental datasets and to the treatment of correlated systematic errors, as well as the quantification of procedural and theoretical uncertainties. 
The extracted $F_2^n$ dataset is utilized to evaluate the $Q^2$ dependence of the Gottfried sum rule and the nonsinglet $F_2^p - F_2^n$ moments. 
To facilitate replication of our study, as well as for general applications, a comprehensive DIS database including all recent JLab 6 GeV measurements, the extracted $F_2^n$, a modified CTEQ-JLab global PDF fit named \texttt{CJ15nlo\_mod}, and grids with calculated proton, neutron and deuteron DIS structure functions at next-to-leading order, are discussed and made publicly available.
\end{abstract}

\date{\today}

\maketitle

\section{Introduction}
 
Experimental data from deep-inelastic scattering (DIS) of leptons from nucleons have been pivotal in the study of nucleon structure for decades. 
In particular, DIS structure function measurements using charged lepton and neutrino beams have been used to extract parton distribution functions (PDFs), investigate scaling and scaling violation, and are generally the ubiquitous tool for studying Quantum Chromodynamics (QCD).
An experimental difficulty in precision DIS nucleon structure studies is the use of nuclear targets, in particular, employment of deuterium as a surrogate neutron target.
Here, Fermi motion, binding, and other nuclear effects render precision neutron measurements difficult, since these may change the shape of the free nucleon structure function.
Yet, it is important to have experimental results for both protons and neutrons to separately determine $u$- and $d$-quark distributions at large parton momentum fractions, $x$, in order to fully understand the partonic structure of nucleons.

Deuteron nuclear corrections have been a challenge to experiments since the first DIS measurements at the Stanford Linear Accelerator Center (SLAC) in the 1970s~\cite{Bodek:1980ar, Arrington2009, Arrington:2011qt}.
Recently, a new testing ground has been provided for these corrections through the CTEQ-Jefferson~Lab (CJ) global PDF analysis effort \cite{Accardi:2009br, Accardi:2011fa, Owens:2012bv, Accardi:2016qay}, where the neutron structure information extracted from DIS data on protons and deuterons using state of the art nuclear corrections has been assessed simultaneously with other high energy scattering datasets.
The latter include, in particular, large rapidity $W$-boson asymmetries in proton--antiproton collisions at the Tevatron \cite{CDF:2009cjw,D0:2013lql}, which are sensitive to the $d$ to $u$ quark PDF ratio at large momentum fractions, and tagged DIS structure functions from the BONuS experiment~\cite{CLAS:2014jvt} at Jefferson Lab (JLab), which select an effective neutron target. 
From agreement among the different datasets and nuclear versus nucleon targets in this more global context \cite{Accardi:2016muk}, confidence may be placed in the nuclear corrections procedure applied, and constraints put on the nuclear correction model itself.
This effort culminated in the CJ15 simultaneous PDF and nuclear correction model extraction \cite{Accardi:2016qay}.
Global analyses similar in spirit to CJ15 have subsequently been performed by Al\"ekhin, Kulagin and Petti (AKP) \cite{Alekhin:2017fpf, Alekhin:2022tip} and the Jefferson Lab Angular Momentum (JAM) Collaboration~\cite{Cocuzza:2021rfn}, with some important differences that will be discussed further in the next Section.
Deuterium data have thus entered an era where the neutron structure function can be determined more accurately.

In fact, knowledge of both proton and neutron structure as measured by the corresponding structure functions $F_2^p$ and $F_2^n$, respectively, and associated nucleon PDFs at large $x$, is central to a broad range of current scientific endeavors. 
Many present and planned experiments at JLab and other facilities seek to understand how the structure of bound objects such as nucleons can be described in terms of their fundamental partonic constituents. 
For example, the limit as $x \to 1$ is a testing ground for multiple perturbative and nonperturbative QCD predictions for the $x$ dependence of PDFs \cite{Melnitchouk:1995fc, PhysRevD.59.034013, BRODSKY1995197, RevModPhys.82.2991, Roberts:2013mja, PhysRevD.103.054029, Ball2016}. 
In the unpolarized sector, this can be accessed by studying the ratio of $d$-quark to $u$-quark PDFs, which is directly constrained by the ratio $F_2^n/F_2^p$ of neutron to proton structure functions.

From a different perspective, reliable knowledge of PDFs at large $x$ is important for searches for new physics signals in collider experiments \cite{Hammou:2023heg}.
This is especially true if the search involves a region where the rapidity of the measured final state is large or the produced mass is large, and the reaction is  sensitive to convolutions of two PDFs evaluated with one value of $x$ being small and the other one large \cite{Brady:2011hb, Accardi:2016ndt}.
Similarly, in searches for novel high-mass resonances \cite{CidVidal:2018eel}, both partons may be evaluated at large $x$.
This issue has been discussed, for example, in the context of the forward-backward asymmetry for large mass dilepton states in processes with Drell-Yan type kinematics \cite{Ball:2022qtp}. 
There it was pointed out that such processes can be sensitive to PDFs evaluated at large values of $x$.
For some PDFs the relevant region of $x$ can be beyond the region where they are constrained by data. 
This then involves extrapolations which can depend critically on the parametrization used.
In such cases disagreement between data and predictions may be a result of insufficient knowledge of the PDFs rather than new physics.
Furthermore, uncertainties in PDFs at large $x$ and low four-momentum transfer squared, $Q^2$, percolate through QCD evolution to affect cross sections at smaller values of the Bjorken-$x$ scaling variable (which at leading order in the strong coupling coincides with the parton momentum fraction $x$) and larger $Q^2$.
In fact, nuclear corrections applied to DIS on deuteron targets, which are most prominent in the valence region at large~$x$, also impact sea quarks and gluon PDFs over a wider range of~$x$, and the achievement of precision in tests of the Standard Model in the electroweak sector will partly depend upon the successful treatment of nuclear corrections~\cite{Accardi:2021ysh}.

From yet another viewpoint, better control of nuclear corrections and a precise knowledge of the $F_2^n/F_2^p$ ratio at large $x$ for non-isoscalarity (the proton and neutron number imbalance) corrections will have a direct, measurable impact on the interpretation of current and future neutrino experiments~\cite{SajjadAthar:2020nvy}, such as NO$\nu$A~\cite{Catano-Mur:2022kyq}, DUNE~\cite{DUNE:2022aul}, and MINER$\nu$A~\cite{MINERvA:2022mnw},
as a large part of the theoretical uncertainty is from the lack of knowledge of the neutrino--nucleus interaction in the large-$x$ valence regime.

On the theoretical front, lattice QCD simulations are providing increasingly precise predictions for moments of PDFs~\cite{Lin:2017snn}, as well as first results for the $x$ dependence of PDFs~\cite{Constantinou:2020hdm}, along with structure functions themselves~\cite{Can:2020sxc, QCDSF:2022btn}, particularly in the valence quark regime.
While calculations can nowadays be performed at physical quark masses, control over a range of systematic uncertainties, such as lattice discretization and volume, is still being improved.
An important benchmark for these calculations is provided by the first moment of the isovector $u-d$ quark PDF, which can be readily compared to moments calculated with PDFs extracted in global QCD analyses.
It is also interesting to more directly compare the lattice moments to experimental data, which can be accomplished by measuring moments of the isovector $F_2^p-F_2^n$ structure functions, and using the operator product expansion to remove target mass corrections (TMCs). 
Phenomenological efforts in this direction have been restricted so far to specific experimental data choices~\cite{Albayrak2018, Kotlorz2021}, and the full power of the global DIS dataset has not yet been leveraged.
Similarly to the $p-n$ moments, neutron structure function data come into play in the experimental determination of the Gottfried sum rule (GSR)~\cite{Gottfried:1967kk, Kataev:2003xp}, through which constraints on the light antiquark distributions in the nucleon may be estimated as a complement to recent data from the SeaQuest experiment at Fermilab~\cite{SeaQuest:2021zxb}. 
Here, again, previous extractions of the GSR have been limited to the analysis of specific datasets~\cite{NewMuon:1993oys, Abbate:2005ct, Kotlorz2021}, and the statistical power of the global DIS dataset has not been fully exploited.

Historically, extractions of the neutron structure function have relied on inclusive lepton-proton and lepton-deuteron DIS data, using models for the nuclear wave function and an iterative extraction prescription~\cite{Bodek:1979rx, Bodek:1980ar}.
Typically, these assume that the deuteron structure function can be written as a convolution of a smearing function (or light-cone momentum distribution of nucleons in the deuterium nucleus), constructed from a nonrelativistic deuteron wave function, and the free nucleon structure functions~\cite{West:1971oyz, Atwood:1972zp}.
Relativistic and nucleon off-shell corrections, which generally cannot be expressed in convolution form~\cite{Melnitchouk:1994rv, Melnitchouk:1993nk}, were included in subsequent analyses \cite{Melnitchouk:1995fc} of SLAC data on protons and deuterons, illustrating the impact of nuclear corrections in the deuteron on the extraction of the neutron structure function at large values of $x$.
Other types of additive corrections to the convolution approximation, such as nuclear shadowing and meson exchange currents, as well as limitations of smearing factor methods, were considered in Ref.~\cite{Umnikov:1994id}.
A rather different approach \cite{Frankfurt:1988nt} assumes scaling with nuclear density in which the nuclear effects in the deuteron are extrapolated from those in heavy nuclei such as $^{56}$Fe~\cite{Whitlow1992, Yang:1998zb}.
Problems with application of the nuclear density model to light nuclei, including ambiguities in defining a nuclear density for deuterium, were discussed in Ref.~\cite{Melnitchouk:1999un}.

In the context of global QCD analysis, nuclear corrections in the deuteron have been extensively studied by the CJ Collaboration~\cite{Accardi:2009br, Accardi:2011fa, Owens:2012bv}, exploring different nuclear wave functions and models of nucleon off-shell corrections.
While earlier CJ analyses relied on model calculations for the off-shell effects, more recent analyses~\cite{Accardi:2016qay} simultaneously fitted the PDF parameters, for given deuteron wave functions, together with the off-shell parameters.
A similar approach was also adopted by Al\"{e}khin {\it et al.}~\cite{Alekhin:2017fpf, Alekhin:2022tip}.
Instead of using deuteron wave functions, Martin {\it et al.}~\cite{Martin:2012da} fitted the entire nuclear effect in the deuteron, including nuclear smearing, phenomenologically in terms of a suitable parametrization describing the deuteron corrections in terms 4 free parameters.
Interestingly, the required deuteron corrections were of the expected general form with a large positive correction at very large $x$ and a dip for $x \sim 0.5$.
Most recently, Ball {\it et al.}~\cite{Ball:2020xqw} proposed an approach agnostic to the nuclear dynamics where the deuteron corrections were treated as an independent source of uncertainty in the determination of the proton PDFs.

In contrast to the PDF-level analyses, Arrington {\it et al.}~\cite{Arrington2009, Arrington:2011qt} performed a study of the neutron structure function entirely at the structure function level, without any reference to partons or QCD.
Surveying various models of nuclear corrections in the literature, including the dependence of the extraction procedure, they assessed the uncertainties from the deuteron wave function, off-shell effects, and different nuclear smearing models used to compute the nuclear corrections, correlating deuterium data with data on heavier nuclei.

The goal of the present work is to apply the CJ deuteron corrections~\cite{Accardi:2016qay} to the large global DIS dataset, including measurements on proton and deuteron targets and their ratios, and to provide an extensive  neutron dataset.
The extraction of the neutron structure function will be data driven, as much as possible, and the model dependence of the procedure confined to the deuteron to free nucleon ratio, 
\begin{equation}
R_{d/N}(x,Q^2) = \frac{F_2^d\,(x,Q^2)}{F_2^N(x,Q^2)},
\label{eq:RdN}
\end{equation}
where the nucleon structure function, $F_2^N = \frac12 (F_2^p + F_2^n)$, and the per-nucleon structure function of the deuteron, $F_2^d$, will be calculated with the PDFs and nuclear correction model simultaneously fitted in the CJ15 next-to-leading order (NLO) global analysis~\cite{Accardi:2016qay}.
Special attention will be devoted to the normalization of the experimental data, and the treatment of correlated systematic errors, as well the quantification of procedural uncertainties.

The obtained neutron, $F_2^n$, and neutron-to-proton ratio, $F_2^{n/p}$, datasets provide a data level representation of the CJ15 nuclear correction model, which may then be used for comparison to other current and future experimental analyses, and as a convenient way to address deuteron nuclear corrections in other PDF extractions or phenomenological analyses.
In this paper, the obtained neutron dataset will be utilized in 
a new data-driven extraction of the GSR, for which we will provide for the first time the $Q^2$ dependence, and an extraction of nonsinglet structure function moments for comparison with recent lattice QCD calculations. 
We will also discuss neutron excess corrections in DIS on heavy nuclear targets.
Of course, other PDF sets, including ones fitted with even higher orders perturbative corrections, could in principle be used instead of the CJ15 PDFs; however, since our ultimate goal is to determine the neutron structure function and not PDFs, any set of PDFs can be used as an effective interpolation grid for the data, as long as it gives a reliable representation of the data and uses an internally consistent set of nuclear corrections.

The extraction of a neutron structure function dataset is particularly timely given that an increasing number of experiments at JLab and elsewhere are or will soon be providing additional data using a variety of experimental techniques to remove or minimize the need for theoretical deuteron nuclear corrections. 
The BONuS12 experiment~\cite{bonus12} at JLab, for instance, uses a spectator tagging technique pioneered in the 6~GeV era to create an effective, essentially free, neutron target~\cite{Baillie:2011za, CLAS:2014jvt}. 
The recently published $F_2^n/F_2^p$ from the MARATHON experiment~\cite{JeffersonLabHallATritium:2021usd}, instead, used a theoretical super-ratio of the EMC effects in the $^3$H and $^3$He mirror nuclei from the AKP nuclear model~\cite{Alekhin:2017fpf, Alekhin:2022tip} to extract $F_2^n/F_2^p$ with reduced nuclear correction uncertainties.
A more general analysis by the JAM Collaboration~\cite{Cocuzza:2021rfn} fitted simultaneously both the nucleon PDFs and the nuclear (off-shell) effects in the $A=3$ nuclei, which, in contrast to the AKP analysis, accounted for differences in the off-shell effects in the proton and neutron.
Nevertheless, having determined the off-shell corrections from a global fit, and the EMC effect in $^3$H and $^3$He, one can use these corrections to extract from the MARATHON data the free neutron structure function.
Care should be taken, however, to ensure that correlations between nucleon off-shell and higher twist (HT) effects and the relevant PDF parameters are properly taken into account~\cite{Alekhin:2022uwc}.

In addition to these experiments that are directly sensitive to the neutron structure function, precise data on inclusive deuteron and proton $F_2$ structure functions are expected from the JLab E12-10-002 experiment~\cite{hallcF2prop}, while the future SoLID program~\cite{solid} will measure parity-violating electron scattering, which is sensitive to the $d/u$ ratio using only a proton target. More proton and neutron tagging and weak current DIS structure function measurements are discussed at the Electron-Ion Collider~\cite{eic-yellow}. 
Moreover, \mbox{$W$-asymmetry} measurements from RHIC at BNL \cite{STAR:2020vuq} and from the LHCb experiment at CERN \cite{CMS:2013pzl, CMS:2012ivw, ATLAS:2011qdp, LHCb:2015okr, CMS:2016qqr} are also sensitive to the $d$ quark, and will thus further help constrain the neutron structure function when included in a global QCD analysis.

The rest of this paper is organized as follows.
In Sec.~\ref{sec:deut_corr} we discuss how deuteron corrections are studied and constrained in the context of global QCD analyses, with particular attention to the CJ15 analysis that will be leveraged in the extraction of the neutron $F_2$ dataset. 
We also analyze differences with the recent AKP and JAM analyses, and briefly discuss their possible sources.
In Sec.~\ref{sec.shujie_data_extraction} we present our data-driven neutron extraction strategy and results, with a careful emphasis on experimental and theoretical uncertainties. 
The data selection criteria and cross-normalization procedure, which is of paramount importance to obtain precise neutron extraction, is also discussed in detail.
In Sec.~\ref{sec.applications} we provide several applications of the neutron data obtained in our analysis, including a new extraction of the Gottfried sum with the first-ever evaluation of its $Q^2$ dependence, and a new extraction of nonsinglet moments for comparison to lattice QCD calculations.
We also provide a new extraction of the $F_2^n/F_2^p$ ratio for use, for example, in neutron excess corrections. 
In Sec.~\ref{sec.conclusions} we summarize our results.

A significant effort has also been made to construct a public database of world DIS data on protons and deuterons, whose usefulness goes beyond the application presented in this paper, which is supplemented by the extracted neutron structure function data and by the bin-centered data used in our phenomenological studies (Appendix~\ref{app:CJdatabase}). 
We also make public the modified set of CJ15 PDF named \texttt{CJ15nlo\_mod} used in our analysis, along with structure function grids in LHAPDF format for proton, neutron and deuteron targets calculated at NLO (Appendix~\ref{app:F2grids}). We provide instructions for using the DIS structure function grids to correct for nuclear (isoscalar) effects in nuclear cross sections in Appendix~\ref{app:isoscalar}.

\section{Deuteron corrections in global QCD analysis}
\label{sec:deut_corr}

Parton flavor separation can be most robustly performed through global QCD analysis~\cite{Jimenez-Delgado:2013sma, Gao:2017yyd, Kovarik:2019xvh, Ethier:2020way}, which  utilizes large collections of data from high-energy collisions sensitive to different underlying PDF combinations. 
Conversely, the extracted PDFs can be utilized in a perturbative pQCD calculation of the unmeasured free neutron $F_2^n$ structure function, or, as we will exploit in this article, of the deuteron to isoscalar nucleon structure function ratio, $R_{d/N}$.
This program necessitates, however, knowledge of the effects on PDFs of the nucleon binding forces in the deuteron in order to properly extract parton-level information from experiments on deuteron targets \cite{Accardi:2016muk}.

Interest in deuteron corrections has recently increased in the global QCD analysis community because these not only impact the $d/u$ ratio at large momentum fractions, but also have important secondary effects on, {\it e.g.}, the gluon or sea-quark PDFs at smaller $x$ values~\cite{Accardi:2021ysh}.
In a global fit, deuteron corrections can be applied at different levels of sophistication. 
For example, one can account for the the additional uncertainty associated with the deuteron effects by iteratively fitting deuteron PDFs in addition to the proton PDFs~\cite{Ball:2020xqw}, parametrize and fit the nuclear deformation of the deuteron structure function~\cite{Martin:2012da, Harland-Lang:2014zoa, Bailey:2020ooq}, or utilize a dynamical model of nuclear interactions to calculate deuteron observables as a double convolution of (off-shell) parton distributions and nucleon wave functions~\cite{Accardi:2016qay, Alekhin:2017fpf, Cocuzza:2021rfn}.
In the CJ approach adopted here, we focus on the latter class of corrections.

The CJ Collaboration \cite{CJ-web} has performed a series of global QCD analyses of unpolarized PDFs \cite{Accardi:2016qay, Owens:2012bv, Accardi:2011fa, Accardi:2009br} with the aim of maximally utilizing DIS data at the highest $x$ values amenable to perturbative QCD analysis.
Special attention has been devoted to deuteron target dynamics, relevant at all energy scales, and to power corrections, such as HTs and TMCs, that become relevant for fixed target experiments probing low values of $Q^2$ and invariant final state mass squared, $W^2$.
To separate the $u$-quark and $d$-quark PDFs, the CJ analysis fits DIS data from both hydrogen and deuterium targets.
The theoretical description of the latter also requires a careful treatment of nuclear interactions, which modify the bound nucleon structure particularly at large values of $x$ at all $Q^2$ scales.
Since the $u$-quark PDF is well constrained by a variety of $ep$, $pp$, and $p \bar p$ scattering data included in the global fit, the $d$-quark PDF extraction is in practice rather sensitive to the neutron structure function, and corrections for nuclear effects in deuterium become a major factor in its accurate determination above $x \approx 0.5$.
This extraction has, therefore, historically suffered from large uncertainties due to the model dependence of these nuclear effects \cite{RevModPhys.82.2991, Roberts:2013mja}.

Beyond deuterium DIS data, the CJ15 PDF analysis~\cite{Accardi:2016qay} studied the impact on the determination of NLO PDFs and their uncertainties from large rapidity charged-lepton~\cite{CDF:2005cgc, D0:2013xqc, D0:2014kma} and $W$-boson asymmetry data~\cite{CDF:2009cjw, D0:2013lql} from proton-antiproton collision at the Fermilab's Tevatron collider.
It also included, for the first time in a global QCD analysis, the novel JLab \bonus data on the free neutron structure function obtained from backward-spectator proton-tagged DIS on a deuterium target~\cite{CLAS:2014jvt} --- a technique utilized to effectively create a free neutron target from deuterium and significantly reduce the nuclear uncertainties that have afflicted previous neutron extractions. 
These two datasets provide the global fit with critical sensitivity to large-$x$ $d$ quarks and neutron structure.  
However, each data type presents positives and negatives within the global QCD fitting context: the $W$-boson asymmetry data exist in the perturbative regime, and are directly sensitive to the $d$ quark without the kinematical smearing that affects the decay lepton measurements; however, the dataset is small and statistical reconstruction of the $W$-boson kinematics is needed to maximize the large-$x$ reach of the measurements.
The tagged DIS data, on the other hand, are more numerous and provide clean neutron data relatively free of deuteron nuclear corrections, but exist in a nonperturbative regime where HT and other effects are of concern. 
The deuteron DIS data, conversely, are rich in kinematic breadth, formed from numerous accurate measurements spanning decades in $x$ and $Q^2$ from multiple laboratories and experiments. 
These data alone would be sufficient to access the neutron structure and $d$-quark PDF --- as can successfully be done to high precision for the $u$ quark with proton data --- were it not for the theoretical nuclear uncertainty associated with extracting neutron structure from deuterium.

These 3 different datasets can then be simultaneously leveraged in the context of global QCD analysis of PDFs to extract the $d$-quark distribution and, at the same time, constrain the deuteron corrections. 
In the CJ15 analysis \cite{Accardi:2016qay}, for example, the nuclear corrections in the deuteron are performed using a smearing formalism in the weak binding approximation~\cite{Kulagin:1994fz, Kulagin:1994cj} that accounts for nuclear Fermi motion and binding through deuteron wave functions calculated in many-body theory, and accounts for bound nucleon deformations of the nucleon PDFs by a Taylor expansion in the nucleon off-shellness, with the expansion coefficient fitted to experimental data. 
The free and bound nucleon structure functions are also corrected for TMC and HT effects.
The interplay of $d$-quark sensitive observables on free protons (such as $W$-boson production in $p \bar p$ collisions) and of DIS data on deuteron targets allows one to disentangle nuclear dynamics and nucleon PDFs in the latter, and to simultaneously constrain both the PDF and off-shell parameters, especially in the large-$x$ region where these are poorly known. 
Tagged deuteron DIS measurements from \bonus then serve to verify the correctness of the deuteron model and, in the future with more statistics expected from the BONuS12 experiment, can put additional constraints on it.

\begin{figure}[t]
\centering
\includegraphics[width=0.9\textwidth]{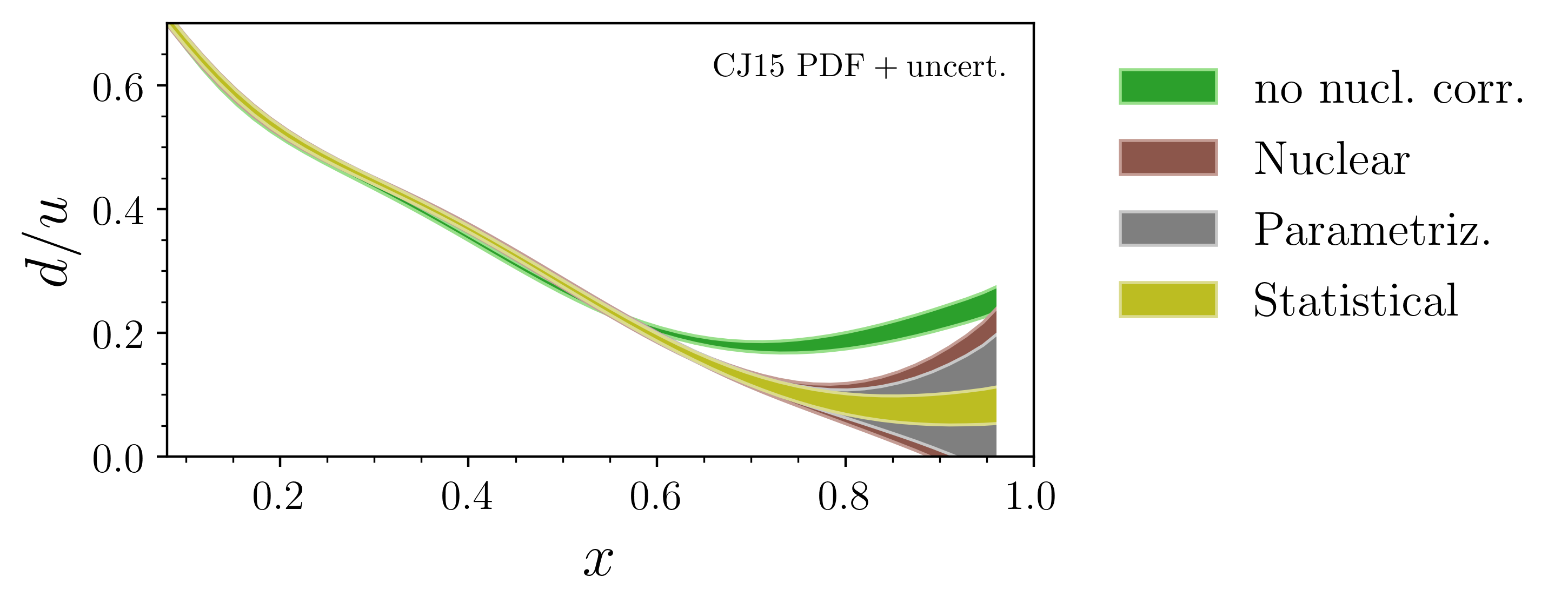}
\caption{\setstretch{1.0} $d/u$ PDF ratio from the CJ15 global QCD analysis~\cite{Accardi:2016qay}. The lower 3 bands show the size of the PDF and off-shell parameter variations compatible with the experimental uncertainties of the fitted experimental data (mustard colored band), as well as the impact of theoretical uncertainties from an extended $d$-quark parametrization (grey band), and the choice of different deuteron wave functions in the calculation of the deuteron structure function (maroon band). The upper green band, with only statistical uncertainties displayed, shows the result of a fit where nuclear corrections are neglected entirely; this fit is strongly disfavored over the nuclear corrected CJ15 fit, with a $\chi^2$ increase of 262 units, largely driven by tensions between deuteron DIS and D\O\ $W$ asymmetry data.} 
\label{fig:du-CJ15}
\end{figure}

The agreement of data from these three disparate reactions, simultaneously contributing to, and well described within, the fundamental QCD framework of the CJ15 global PDF fit thus provides one not only with an improved knowledge of the $d$-quark PDF, but also with confidence in the ability to access the neutron using the deuteron nuclear corrections deployed in that global analysis.
If the nuclear corrections were inadequate beyond expected uncertainties, the three data types would exhibit significant tension within the global fitting framework. 
The fact that they do not gives confidence in the applied nuclear corrections and facilitates the neutron dataset extraction provided in this present work, with the global analysis framework enabling the necessary evaluation of associated procedural uncertainties. 
As an example, the $d/u$ PDF ratio obtained in the CJ15 analysis is displayed in Fig.~\ref{fig:du-CJ15}, showing the importance of including nuclear corrections when fitting deuteron target DIS data.
The ``statistical'' PDF uncertainties propagated from the experimental uncertainties are supplemented by an estimate of the theoretical uncertainties due to PDF parametrization choice and the deuteron many-body wave function.

\begin{figure}[t]
  \centering
  \includegraphics[width=0.65\textwidth]{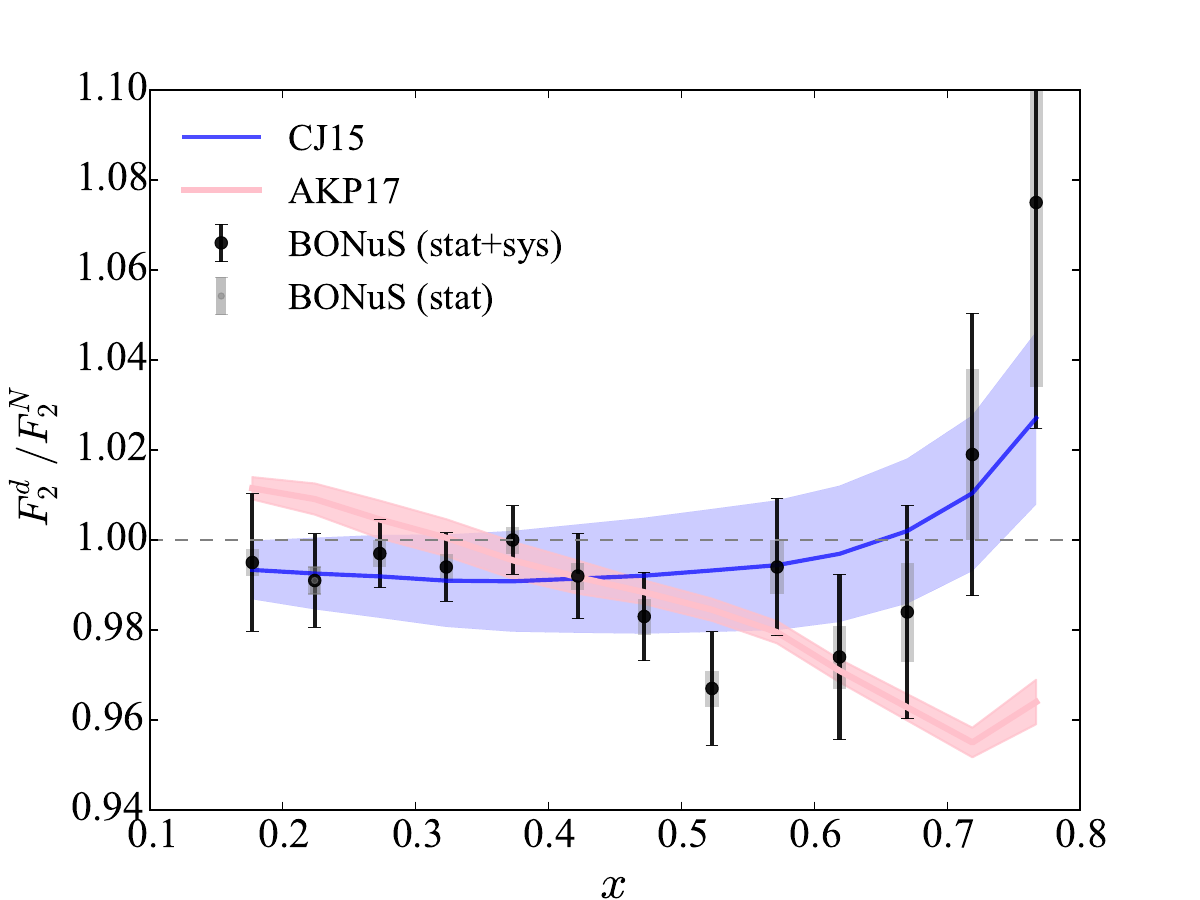}
  \caption{\setstretch{1.0} Deuteron to nucleon $F_2$ structure function ratio from the CJ15 \cite{Accardi:2016qay} and AKP17~\cite{Alekhin:2017fpf} global QCD analyses compared to the \bonus experimental extraction \cite{bonus_emc}. The bands display the fit uncertainty at the 90\% CL (the published 68\% CL AKP band was rescaled by a factor 1.646). Data at $x\gtrsim 0.5$ have an average invariant mass squared $2 < \langle W^2 \rangle < 3 \textrm{\ GeV}^2$, and the theoretical calculations here
  should be understood 
  in the spirit of quark-hadron duality \cite{Melnitchouk:2005zr}.
  } 
\label{fig:dN-CJ-AKP-BoNUS}
\end{figure}

In contrast to the CJ15 analysis, the global fits performed by Al\"{e}khin {\it et al.} in 2017 (AKP17)~\cite{Alekhin:2017fpf} and revisited in 2021 (AKP21)~\cite{Alekhin:2022tip} resulted in marked differences in the fitted off-shell function and the $d/u$ PDF ratio, which is softer than the CJ15 result at $x \gtrsim 0.7$ and (by construction) tends to 0 as $x \to 1$.
At the same time, the AKP17/21 neutron structure function, which is indirectly extracted from the measured proton and deuteron data, also differs from the CJ15 result, especially at large $x$ where it is actually harder than in the CJ15 case.
This can be appreciated from Fig.~\ref{fig:dN-CJ-AKP-BoNUS}, which compares the calculated $R_{d/N}$ ratio to the experimental values obtained by the \bonus Collaboration combining their tagged, quasi-free neutron to deuteron ratio $F_2^n/F_2^d$ measurements with the world data on proton, $F_2^p$, and deuteron, $F_2^d$, structure functions.
The recent JAM21 global analysis~\cite{Cocuzza:2021rfn}, on the other hand, which considered a dataset similar to those used in the CJ15 and AKP17/21 analyses, along with the $^3$He/$^3$H cross section ratio from the MARATHON experiment~\cite{JeffersonLabHallATritium:2021usd}, finds the $d/u$, $F_2^n/F_2^p$, and $R_{d/N}$ ratios that are similar to the present results.
The extracted ratios and the isospin-averaged off-shell functions were also consistent with the results of the CJ15 analysis \cite{Accardi:2016qay}, but were less compatible with the AKP off-shell corrections.

The discrepancy between the deuteron to nucleon ratio $R_{d/N}$ calculated in independent global QCD analyses can have multiple sources; among these are the choice of fitted datasets and PDF parametrization, as well as the phenomenological implementation of HT effects, TMCs and nuclear corrections~\cite{Li-Paris-2021, Alekhin:2022tip, Alekhin:2022uwc}. 
Tracking down and disentangling from each other their effects is a complex task since they combine nonlinearly.
In fact, HT corrections and their interplay with off-shell effects might be of particular relevance to the extraction of the neutron structure function~\cite{Li-Paris-2021}.
That this may be the case can be appreciated by noticing that the AKP $d$-quark PDF is softer than the corresponding CJ15 distribution at large values of $x$, so that larger HTs are needed to bring the AKP $R_{d/N}$ ratio below the CJ15 ratio in Fig.~\ref{fig:dN-CJ-AKP-BoNUS}.

An alternative approach to deuteron corrections that is agnostic to the nuclear dynamics was recently proposed by Ball, Nocera and Pearson \cite{Ball:2020xqw}, where these are treated as an independent source of systematic error on the determination of the proton PDFs. 
In their approach, deuteron PDFs are separately fitted to deuteron target data, and the uncertainties on the deuteron PDFs utilized to deweight deuteron data when utilizing these alongside the rest of the global data set in a proton PDF fit. 
In contrast, in the approach followed by the CJ, AKP and JAM collaborations the deuteron wave function is utilized to calculate Fermi motion and binding effects on the nucleons in the deuteron, and off-shell PDF deformations are constrained by the global dataset without the need of a separate fit of deuteron PDFs. 
The $R_{d/N}$ ratio found in Fig.~7 of Ref.~\cite{Ball:2020xqw} is nonetheless very similar to the CJ15 ratio displayed in Fig.~\ref{fig:dN-CJ-AKP-BoNUS}, as well as to the $R_{d/N}$ ratio fitted alongside the proton PDFs to the global data in the MMHT2014 analysis \cite{Harland-Lang:2014zoa}.

A comprehensive investigation of the role of various theoretical corrections is outside of the scope of this article, but remains an important objective for future studies. 
These will be facilitated by the development of new statistical analysis methods, such as the recently developed $L_2$ sensitivity method~\cite{Hobbs:2019gob}, which was utilized in Ref.~\cite{Accardi:2021ysh} to distinguish the data-driven features of a global fit from methodological differences in the CJ15 and CT18 PDF global analyses.

\section{Neutron Structure Function Extraction} 
\label{sec.shujie_data_extraction}

The central aim of this paper is to perform a data-driven extraction of a cross-normalized set of neutron $F_2^n$ structure functions from a set of experiments that measured $F_2$ on both proton and deuteron targets.
A secondary aim is to extract neutron to proton $F_2^n/F_2^p$ ratios from deuteron to proton ratio data, as well as from spectator tagging experiments, such as BONuS, that provide neutron to deuteron ratio measurements.

This program can be accomplished with minimal sensitivity to the theoretical inputs by systematically applying to the data the theoretically calculated correction ratio $R_{d/N}^{\rm th}(x,Q^2)$, as in Eq.~(\ref{eq:RdN}), across a range of $x$ and $Q^2$,
effectively converting the measured deuteron $F_2^d$ structure function (measured alone or as part of a $d/p$ or tagged $n/d$ structure function ratio) into a superposition of free proton and free neutron structure functions.
The neutron $F_2^n$ structure function can then be extracted by suitably subtracting the measured free proton contribution from the observable under consideration. 
For consistency, the same theoretical calculations utilized for $R_{d/N}^{\rm th}$ 
will also be used to cross-normalize data from different experiments and from different targets within one experiment in order to control relative fluctuations in their systematic shifts.

In the present analysis we perform the needed theoretical calculations utilizing the PDFs and deuteron correction model simultaneously fitted in the CJ15 global QCD analysis~\cite{Accardi:2016qay}.
The choice of CJ15 as a model to calculate the $F_2$ structure functions of free nucleons and to evaluate the nuclear effects in the deuteron affects the neutron data extraction in two places: firstly, the cross-normalization of the experimental proton and deuteron data, and secondly the calculation of the nuclear correction ratio, $R_{d/N}^{\rm th}$.
As mentioned earlier, a global analysis similar in spirit to CJ15 was performed by AKP in Refs.~\cite{Alekhin:2017fpf, Alekhin:2022tip}, but resulted in marked differences in the $d/u$ PDF ratio and, more importantly for this analysis, in the obtained $R_{d/N}^{\rm th}$ ratio.
These differences would result in a systematic shift of the extracted neutron data if the AKP model were to be adopted instead of the CJ15 model.
However, a full evaluation of this systematic effects would require use of the AKP fitting framework, and goes beyond the scope of this paper.

\subsection{Proton and deuteron data selection and cross-normalization}
\label{sec:pd_cross_normalization}

In order to provide a complete neutron dataset over the widest possible kinematic range, we extended the DIS database utilized in the CJ15 global QCD analysis \cite{Accardi:2016qay} to include all the relevant inclusive measurements from the JLab~6~GeV experimental program.
We also revisited correlated errors in the existing CJ15 $F_2$ data collection (such as in the NMC~\cite{Arneodo1995} and SLAC~\cite{Whitlow1990} datasets), and included data from the SLAC-140x~\cite{e140x} experiment.
This extended and up-to-date DIS database 
is discussed further in Appendix~\ref{app:CJdatabase}.

\begin{figure}[p]
  \centering
  \includegraphics[width=0.95\textwidth]{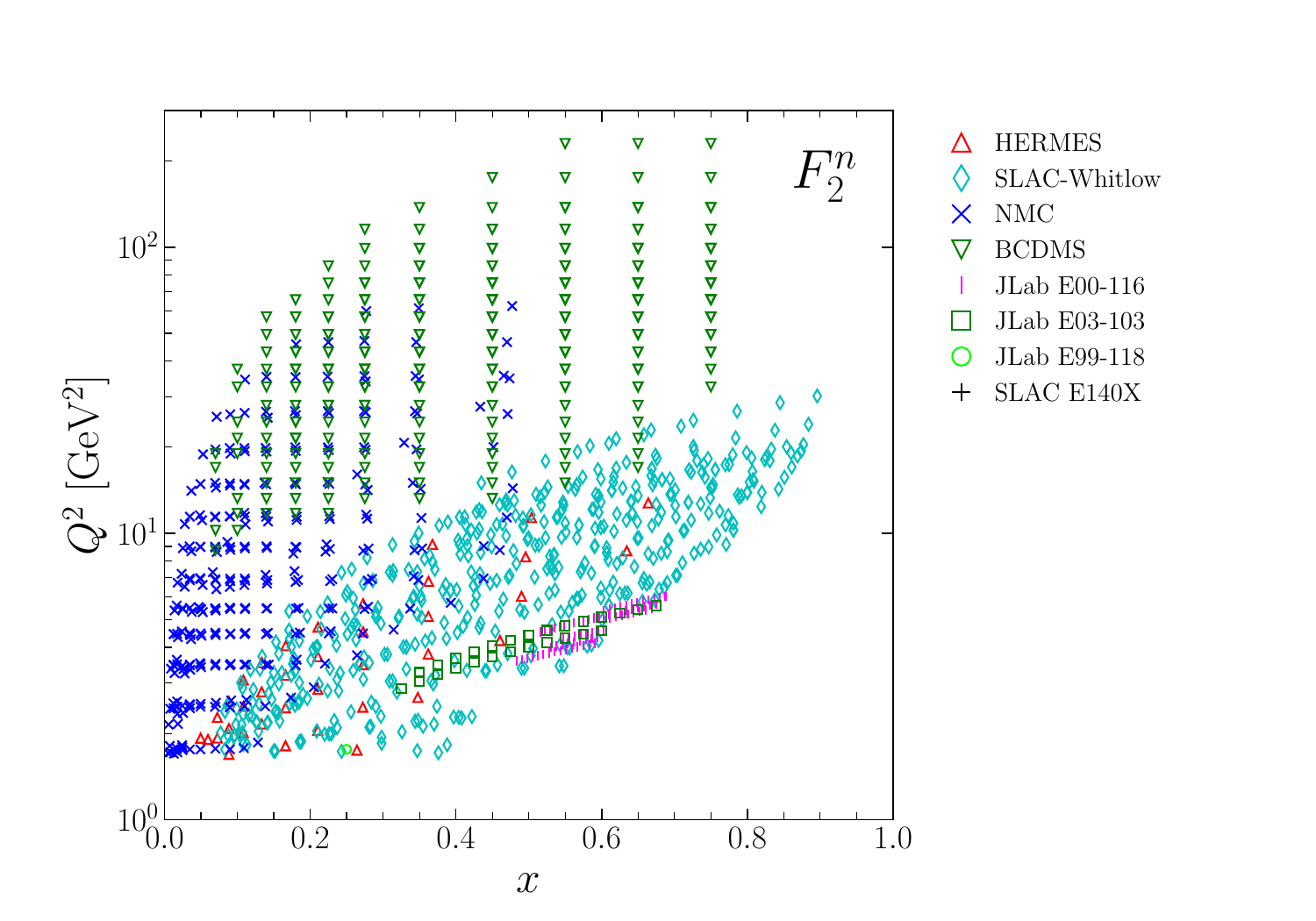}\vspace*{-0.3cm}
  \includegraphics[width=0.95\textwidth]{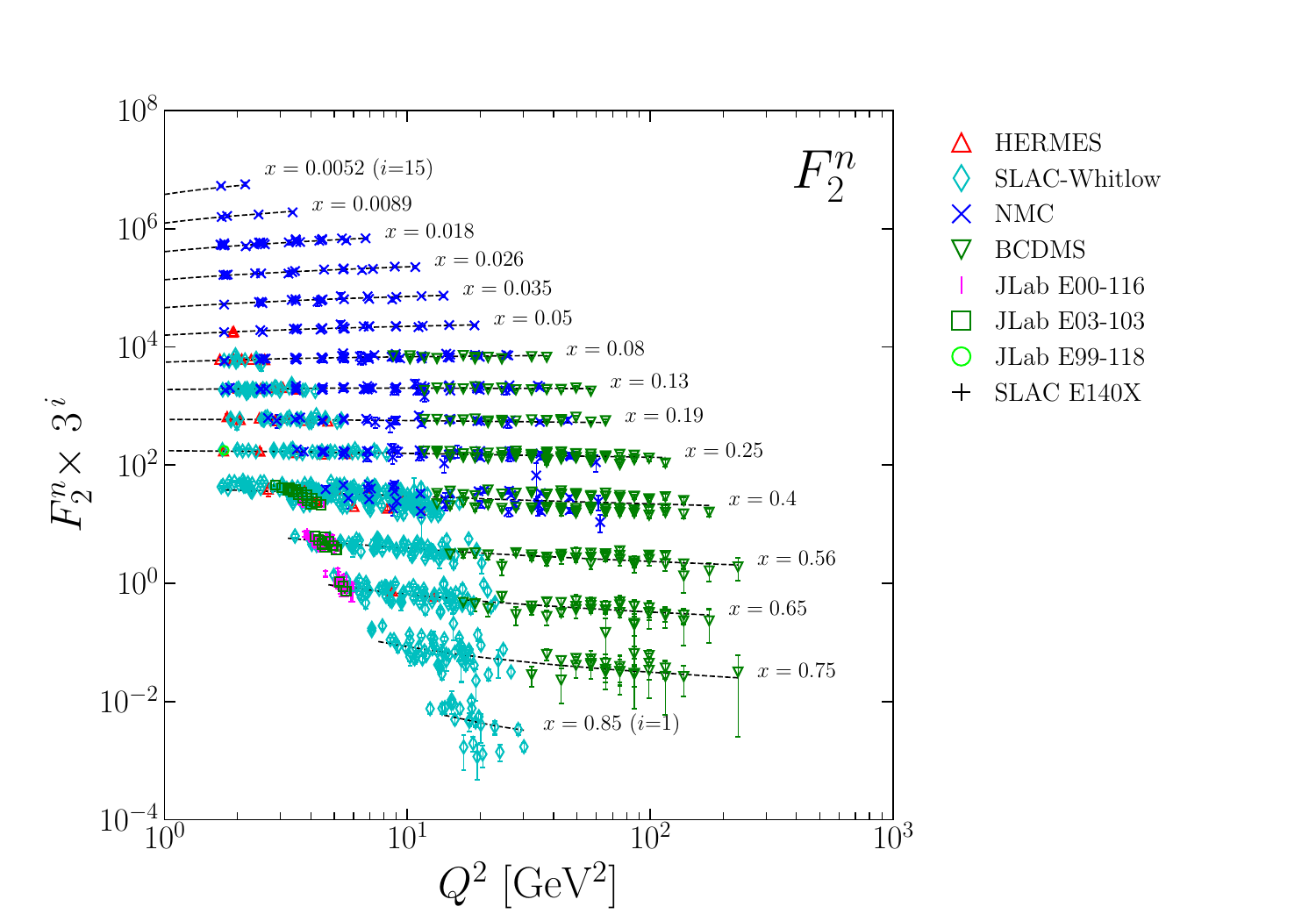}\vspace*{-0.2cm}
  \caption{\setstretch{1.0} ``Matched'' kinematic range (upper panel) and $F_2$ value binned in $x$ (lower panel) of the  neutron structure functions extracted from world data (see also Table~\ref{tab:DIS-matched-data}). The $F_2$ values are cut at $W^2=3.5$~GeV$^2$. 
  }
  \label{fig:F2n-kin}
\end{figure}

\begin{figure}[p]
  \centering
  \includegraphics[width=0.95\textwidth]{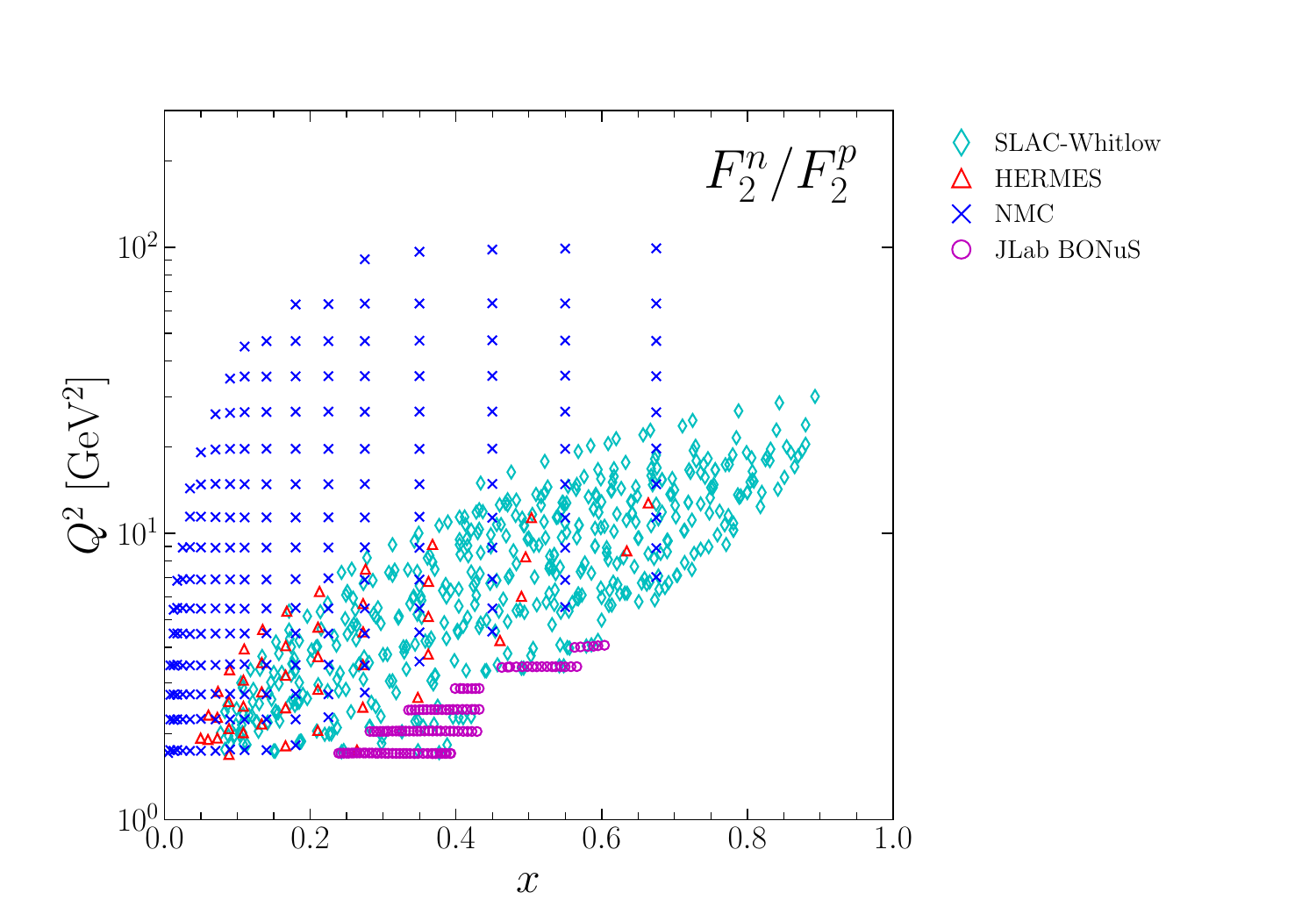}\vspace*{-0.3cm}
  \includegraphics[width=0.95\textwidth]{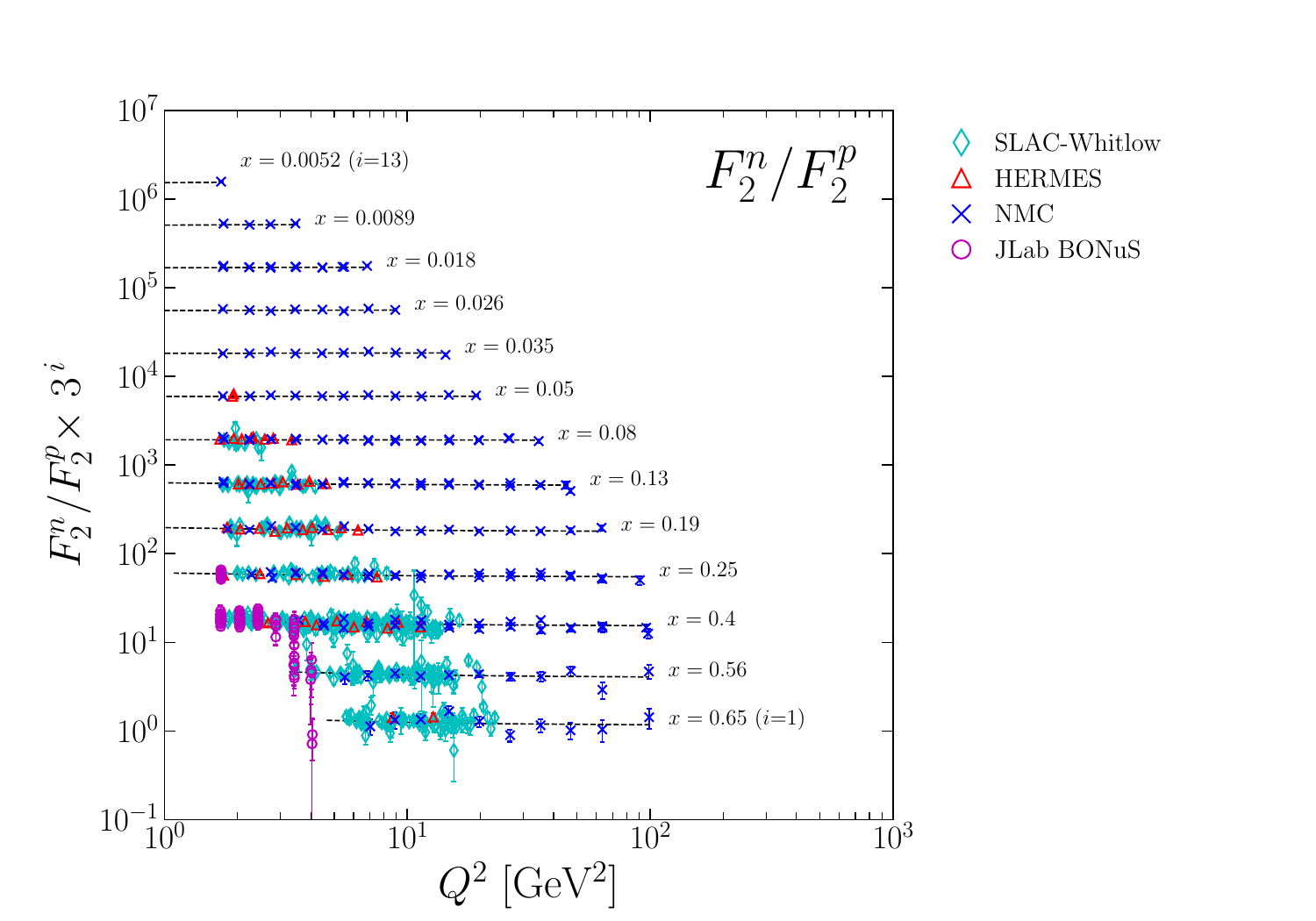}\vspace*{-0.2cm}
  \caption{\setstretch{1.0} Kinematic range (upper panel) and $F_2$ value binned in $x$ (lower panel) of the  $n/p$ ratio extracted from world data (see also Table~\ref{tab:DIS-ratio-data}). The $F_2$ ratios are cut at $W^2=3.5$~GeV$^2$.}
  \label{fig:F2np-kin}
\end{figure}

Our extraction of the neutron structure function $F_2^n$ (see Sec.~\ref{sec:F2n_extraction} below) requires us to simultaneously manipulate proton and deuteron structure functions measured at the same values of $x$ and $Q^2$ in order to minimize the size of systematic uncertainties.
To this purpose, we select pairs of proton and deuteron data according to the following criteria:
\begin{enumerate}
  \item proton and deuteron $F_2$ 
  data were measured within the same experiment and at the same beam energy;
  \item the corresponding Bjorken-$x$ values agree to within an interval $\Delta x=0.01$;
  \item the $Q^2$ values agree to within 1\%.
\end{enumerate} 
To be consistent with the CJ15 analysis~\cite{Accardi:2016qay}, we also require that all selected data satisfy the cut $Q^2 > 1.69$~GeV$^2$, which marks the starting scale for QCD evolution of the PDFs, and $W^2 > 3.5$~GeV$^2$ to select DIS events.
Note that we increased the $W^2$ cut from 3~GeV$^2$ (used in the CJ15 analysis) to 3.5~GeV$^2$ to better reject the resonance contamination in the large-$x$ and low-$Q^2$ data.
Overall, we obtain 1192 matched proton and deuteron $F_2$ data points covering the range $x=0.005 - 0.896$ and $Q^2 = 1.69 - 230$~GeV$^2$, as shown in
Fig.~\ref{fig:F2n-kin} and summarized in Table~\ref{tab:DIS-matched-data}.
Conversely, no matching of $d/p$ and tagged $n/d$ data is required to extract the neutron-to-proton structure function ratio,
\begin{eqnarray}
    R_{n/p}(x,Q^2) &\equiv& \frac{F_2^n(x,Q^2)}{F_2^p(x,Q^2)}.
\label{eq.F2np}
\end{eqnarray}
The available ratio data, listed in Table~\ref{tab:DIS-ratio-data}, have a smaller multiplicity but also smaller statistical and systematic uncertainties. 
Their kinematic coverage is illustrated in Fig.~\ref{fig:F2np-kin}.

\begin{table}[hp]
\centering
\caption{\setstretch{1.0} Matched proton and deuteron structure function datasets used in the current $F_2^n$ extraction. The number $N_\text{match}$ of matched data points is listed, along with the experimental normalization uncertainties, $\delta n$. The nuisance parameter $\lambda^{\rm norm}$ is needed to normalize the data to the CJ15-based theoretical calculations, and the size of its procedural uncertainty  $\delta^\CJ\lambda$ is discussed in Sec.~\ref{sec:uncertainties}. Datasets marked with ``${\, *\,}$'' have previously been included in the CJ15 analysis (except the NMC $d/p$ ratio was used instead of the $d$ structure function in the PDF fit). For consistency with the CJ15 analysis~\cite{Accardi:2016qay}, 8 data points from HERMES have been excluded in this work. The E99-118 data come with no normalization uncertainties, and thus no $\lambda^{\rm norm}$ are reported, although we note that a normalization shift of 1.55(37)\% and 2.41(50)\% would bring the $p$ and $d$ datasets, respectively, to good agreements with CJ15. \\}
\begin{tabular}{l|r|r|r}    
    \hline
    \thead{$F_2^{p,d}$ experiment} & 
    \thead{\begin{tabular}[c]{@{}c@{}} ~$N_\text{match}$~ \end{tabular}} & \thead{\begin{tabular}[c]{@{}c@{}} $\delta n_{p,d}$ (\%) \end{tabular} } & \thead{$\lambda^{\rm norm}_{p,d}$}
\\ \hline
    ~*SLAC Whitlow \cite{Whitlow1990}       
    & 478~  &  2.1,~  1.7~  
    & $0.38(14)$,	~~$0.65(18)$~
\\ 
    ~~\,SLAC E140x \cite{e140x}         
    & 9~    &  ~1.73,\!  1.73~ 
    & $-0.12(6)$,~	$-0.32(9)$~~\,	
\\  
    ~*NMC \cite{Arneodo1997a}
    & 275~  &  2.5,~  2.5~  
    & $0.40(16)$,~	~$0.04(12)$~
\\ 
    ~*BCDMS \cite{Benvenuti1989, Benvenuti1990}   
    & 254~  &  3.0,~  3.0~  
    & $-0.90(7)$,~	$-0.23(7)$~~\,	
\\ 
    ~*HERMES \cite{hermes}      
    & 37~   &  7.5,~  7.5~  
    & $-0.03(3)$,~	$-0.27(4)$~~\,
\\ 
    ~~\,JLab E99-118 \cite{e99118_f2}      
    & 2~    &  &  
\\ 
    ~~\,JLab E03-103 \cite{e03103, 03103thesis}~ 
    & 32~   &  2.5,~  2.5~
    & $-0.6(4)$,~~\,   $-0.88(16)$~	
\\ 
    ~*JLab E00-116 \cite{e00116_f2}          
    & 97~   &  1.75,\!  1.75~
    & ~~~$-0.59(17)$,\!	$-1.41(25)$~	
\\ \hline   
\end{tabular}
\label{tab:DIS-matched-data}
\end{table}

\begin{table}[ht]
\centering
\caption{\setstretch{1.0}Structure function ratio data used in the neutron-to-proton structure function ratio extraction. As for Table~\ref{tab:DIS-matched-data}, $\delta n$ is the experimental normalization uncertainty, and $\lambda^{\rm norm}$ is the nuisance parameter needed for cross-normalization through the CJ15 global QCD analysis. Data marked with ``$\, *\, $'' were included in the CJ15 global PDF fit.\\}
\begin{tabular}{l|r|c|l}
    \hline
    \thead{ratio experiment} &\thead{~$N_\text{data}$~~}   
    & \thead{~$\delta n\, (\%)$~~} & \thead{~~$\lambda^{\rm norm}$} 
\\ \hline
    ~*\bonus $F_2^n/F_2^d$ \cite{CLAS:2014jvt}~        
    & 115~    &  4.5    &  ~$-0.36(9)$    
\\ 
    ~*NMC  $F_2^d/F_2^p$ \cite{Arneodo1997}~           
    & 189~    &  0.3    &  ~$-1(1)$    
\\ 
    ~~HERMES $F_2^d/F_2^p$ \cite{hermes}~            
    & ~45~    &  1.4    &  ~$-0.57(21)$~    
\\ 
    ~~SLAC-Whitlow $F_2^d/F_2^p$ \cite{Whitlow1990}~       
    & 487~    &  1.0    &  ~~~\,$0.2(3)$    
\\ \hline   
\end{tabular}
\label{tab:DIS-ratio-data}
\end{table}

\newpage
Each experimental dataset typically provides three groups of uncertainties:
\begin{enumerate}
\item 
A multiplicative overall relative normalization uncertainty $\delta n$. 
This allows all points in a given dataset to be scaled by a common normalization factor, 
\begin{align}
   n = 1+\lambda^{\rm norm}\, \delta n \ ,
\label{eq:norm}
\end{align}
where $\lambda^{\rm norm}$ is a normally distributed stochastic variable with zero mean and unit standard deviation.
\item 
Point-to-point uncorrelated statistical, $\delta_{\rm stat} D_i$, and systematic, $\delta_{\rm syst}D_i$, uncertainties, for the set's $i$-th data point $D_i$.
These are summed in quadrature to obtain the total uncorrelated uncertainty, $\delta D_i$.
\item 
A number $K$ of point-to-point correlated additive systematic uncertainties $\vec\beta_k=(\beta_{k,1},\beta_{k,2},...,\beta_{k,N_{\rm data}})$, with $k=1, \ldots, K$ labeling the uncertainty sources, and $N_{\rm data}$ the number of data points in that dataset.
The $i$-th data point can then be shifted by an amount 
\begin{equation}
    \Delta_{i}= \sum_k \lambda_k\, \beta_{k,i}\, ,
\end{equation}
where $\lambda_k$ are normally distributed stochastic variables with zero mean and unit standard deviation. 
\end{enumerate}
The ``nuisance parameters'' $\lambda^{\rm norm}$ and $\lambda_k$ from each experiment allow data points from proton and deuteron datasets to independently fluctuate within their reported normalization uncertainty, and also, when available, within their correlated uncertainties \cite{Kovarik:2019xvh}.
Note that the NMC experiment also provides information on the cross-correlations between their measurements on proton and on deuteron targets; however, in this analysis we have conservatively treated these two targets as fully uncorrelated, as for all other datasets.

Since we aim at extracting the neutron structure function by subtraction of the proton component from its matched deuteron measurement, it is vital to determine an optimal set of nuisance parameters for all experiments, or in other words to cross-normalize the datasets. 
In particular, cross-normalization of the proton and deuteron data is needed to avoid large fluctuations in the neutron extraction due to their relative systematic shifts. 
To accomplish this we choose to fix $\lambda^{\rm norm}$ and $\lambda_k$ of each experiment by comparing the experimental data $D_i$ to a corresponding theoretical value $T_i$ calculated in perturbative QCD at the measured kinematics of each data point.
The following $\chi^2$ function is then minimized with respect to $\lambda^{\rm norm}$ and the $\lambda_k$ parameters of each experiment (exp),
\begin{equation} 
  \chi^2  = \sum_\text{exp} \left[ 
     \sum_{i=1}^{N_{\rm data}}
     \Bigg( \frac{ D_i + \Delta_i - T_i/n}{\delta D_i} \Bigg)^2
  + \big( \lambda^{\rm norm} \big)^2 
  + \sum_{k=1}^K\lambda_{k}^2 \, \right]_\text{exp} \ .
  \label{eq:chi2_CJ}
\end{equation}
Note that measurements on different targets in a given experiment are here labeled and treated as separate experiments. 
(Only the NMC reported correlations between different target measurements, but we conservatively consider these as statistically independent.)

The theoretical value $T_i$ for each kinematic point is calculated in perturbative QCD at NLO using the PDFs and the deuteron correction model determined in the CJ15 global analysis.
While these are kept fixed, the nuisance parameters $\lambda^{\rm norm}$ and $\lambda_k$ of all experiments are simultaneously fitted. 
(Note that, in fact, the CJ15 PDFs were fitted to a dataset that also included measurements of jet production and Drell-Yan lepton pair production in hadron-hadron collisions, besides the DIS measurements highlighted in Tables~\ref{tab:DIS-matched-data} and \ref{tab:DIS-ratio-data}, thus imposing tighter constraints on the nuisance parameters than allowed by DIS data alone.)
The obtained normalization parameters $\lambda^{\rm norm}$ are listed in Tables~\ref{tab:DIS-matched-data} and \ref{tab:DIS-ratio-data}, alongside their fit uncertainty, to be discussed in Sec.~\ref{sec:uncertainties}. 
Finally, we define the central value of the cross-normalized $F_2$ data, denoted by the superscript ``(0)'', as
\begin{equation}
  \widehat D_i^{(0)} = n^{(0)} \big( D_i + \Delta_i^{(0)} \big)\,
\end{equation}
where 
$n^{(0)}=1+\lambda^{\rm norm\, (0)}\, \delta n$ and $\Delta_i^{(0)}= \sum_k  \lambda_k^{(0)}\,\beta_{k,i}$ are the fitted normalization factors and correlated data shifts.

\begin{figure}[p]
\vskip.5cm
\centering
  \includegraphics[width=14.5cm]{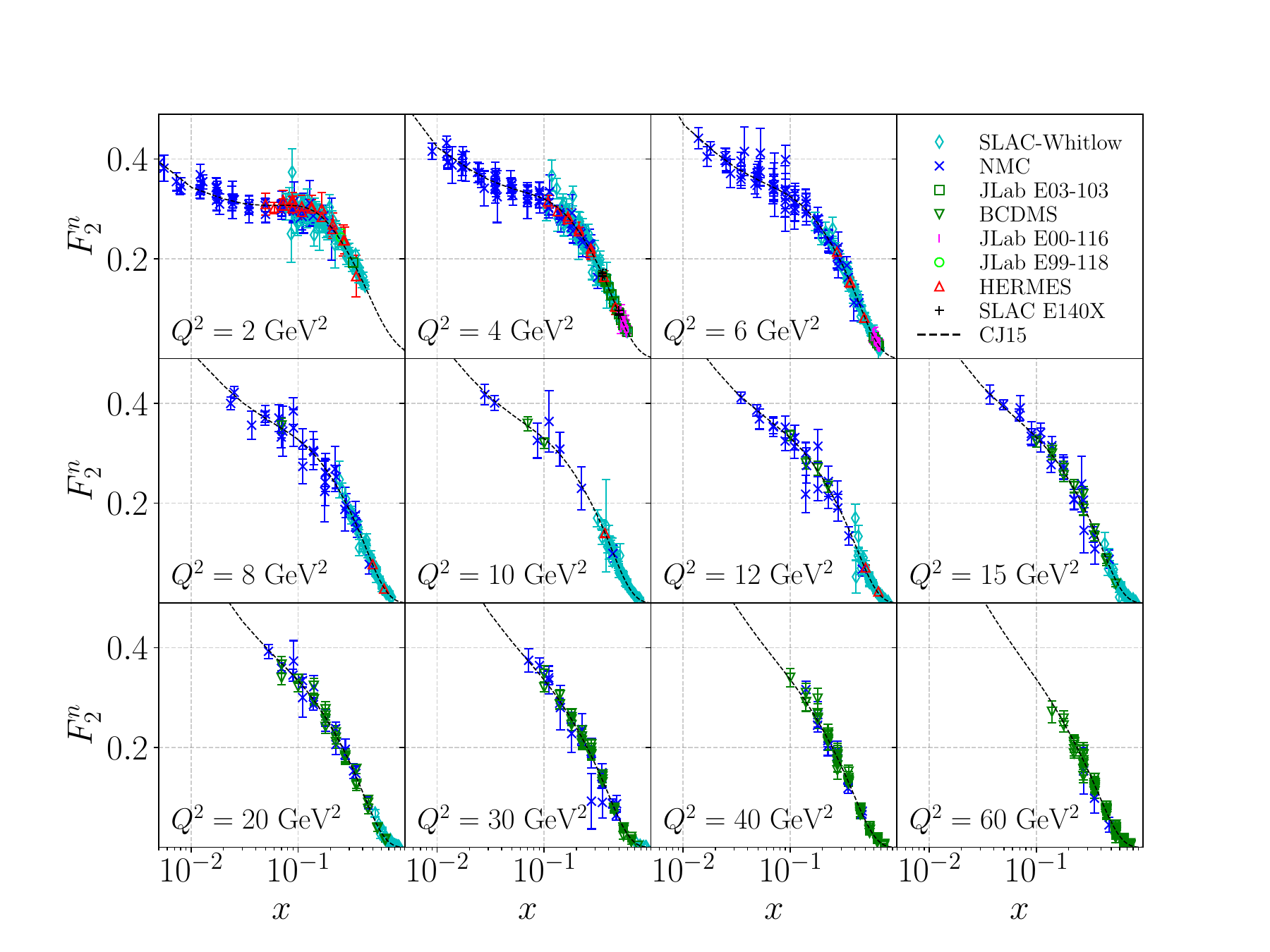}\vspace*{-0.3cm}
  \includegraphics[width=14.5cm]{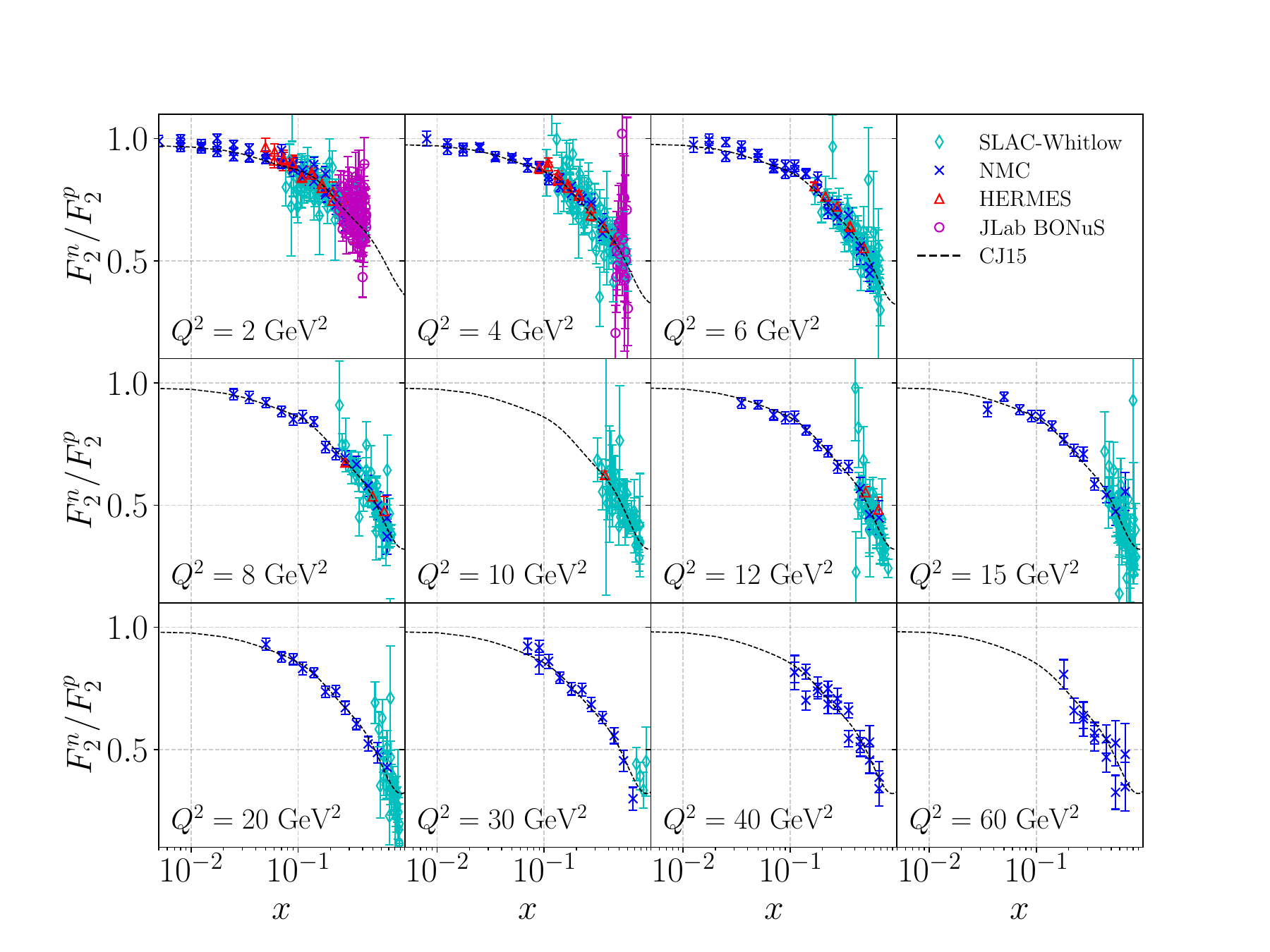}\vspace*{-0.6cm}
  \caption{\setstretch{1.0}
  Extracted $F_2$ neutron structure function (upper) and $n/p$ ratio (lower) as a function of $x$ in selected $Q^2$ bins, compared with the CJ15 fit (dashed black line). The data are bin-centered in $Q^2$ and $x$ for clarity, and the experimental ($\delta^u$) and procedural ($\delta^\CJ$) uncertainties are added in quadrature.}
\label{fig:f2_bincentered} 
\end{figure}

\subsection{Neutron $F_2^n$ and neutron-to-proton ratio extraction}
\label{sec:F2n_extraction}

The cross-normalized $\widehat F_2^p$ and $\widehat F_2^d$ data can now be used to extract the neutron structure function, $\widehat F_2^n$.
As noted above, in order to do this we use the deuteron correction factor 
  $R_{d/N}^\CJ(x,Q^2)$
as in Eq.~(\ref{eq:RdN}) to remove the nuclear effects from the deuteron data, with the proton and deuteron structure functions computed at NLO in perturbative QCD using the PDFs and nuclear correction model from the CJ15 global QCD analysis~\cite{Accardi:2016qay}. 
The central value of the neutron structure function can then be obtained as
\begin{equation}
    \widehat{F}_2^{n(0)}(x,Q^2)
    = \frac{ 2\, \widehat{F}_2^{d(0)}(x,Q^2)_{\rm exp} }{ R_{d/N}^\CJ(x,Q^2) }\,
    -\, \widehat{F}_2^{p(0)}(x,Q^2)_{\rm exp}\, .
  \label{eq:f2n}
\end{equation}
This formula first converts the experimentally measured deuteron data into the sum of free proton and free neutron structure functions, then subtracts from this the experimentally measured proton contribution to obtain the neutron structure function.
An alternative approach defining 
    $F_2^n=F_{2,{\rm exp}}^d\, \big[ F_2^n/F_2^d \big]_\CJ$ 
was discarded because it trades a smaller amount of experimental information for a larger amount of theoretical input with its associated uncertainties.
Conversely, Eq.~\eqref{eq:f2n} minimizes the use of the theoretical model in accordance with our goal of performing a data-oriented neutron extraction. 
The result is displayed in Fig.~\ref{fig:F2n-kin} as a function of $Q^2$ in selected $x$ bins, and in Fig.~\ref{fig:f2_bincentered}  as a function of $x$ in selected $Q^2$ bins. The uncertainties will be discussed in Sec.~\ref{sec:uncertainties} and bin centering will be addressed in Sec.~\ref{sec:bin_centering}.
The $x$ and $Q^2$ kinematic coverage of the extracted data corresponds to that of the matched $p$ and $d$ data, as pictured in Fig.~\ref{fig:F2n-kin}.

\begin{figure}
    \begin{center}
    \includegraphics[width=0.7\textwidth]{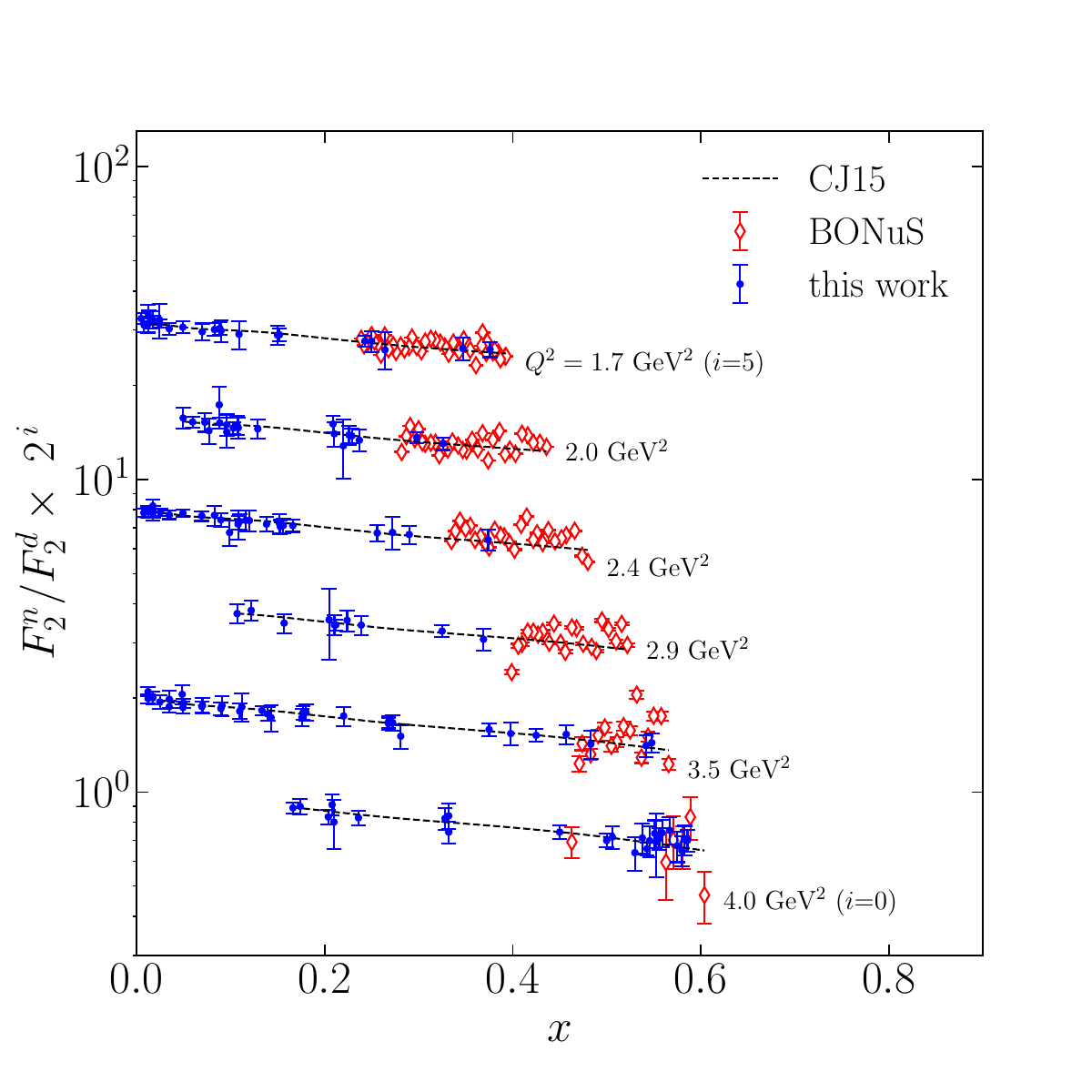}
    \end{center}
    \vspace*{-1cm}
    \caption{
        \setstretch{1.0} Neutron to deuteron ratios $F_2^n/F_2^d$ from this extraction (blue circles) compared with the JLab BONuS data (red diamonds) and the ratio calculated from CJ15 (dashed black lines). The $F_2^n$/$F_2^d$ values extracted in this paper are selected within a 0.1~GeV$^2$ slice of each quoted $Q^2$ value.
    }
    \label{fig:f2nd}
\end{figure}

This procedure cannot be directly benchmarked against experimental data for the neutron structure function. 
However, as a consistency check, we can combine the extracted $F_2^n$ with the measured deuteron structure function, and compare their ratio to the 
    $R_{n/d} = F_2^n/F_2^d$ 
ratio experimentally measured by the 6~GeV BONuS experiment at JLab via spectator proton tagging~\cite{CLAS:2014jvt}.
As shown in Fig.~\ref{fig:f2nd}, the neutron to deuteron ratio extracted as described above agrees with BONuS data, and conversely validates the 6~GeV BONuS analysis.

Analogously to the neutron $F_2^n$ extraction, the central values of the neutron-to-proton ratio $R_{n/p}$ can be obtained from the experimental $d/p$ or $n/d$ ratio data, utilizing the deuteron correction factor from Eq.~(\ref{eq:RdN}) in analogy with the neutron extraction of Eq.~\eqref{eq:f2n},
%
\begin{equation}
  \widehat{R}^{(0)}_{n/p}\, \equiv\, \frac{2\, \widehat{R}_{d/p}^{\rm exp,(0)}}{R_{d/N}^\CJ}-1 \ ,
\end{equation}              
and
%
\begin{equation}
  \widehat{R}^{(0)}_{n/p}\, \equiv\, \frac{ \widehat{R}_{n/d}^{\rm exp,(0)}\ R_{d/N}^\CJ}
  {2-\widehat{R}_{n/d}^{\rm exp,(0)}\ R_{d/N}^\CJ } \ .
\end{equation}
The bin-centered version of the extracted $n/p$ ratio is displayed as a function of $x$ in Fig.~\ref{fig:F2np-kin} and as a function of $Q^2$ in Fig.~\ref{fig:f2_bincentered}.

\subsection{Uncertainties}
\label{sec:uncertainties}

The extracted $\widehat F_2^n$ structure function (and the $\widehat R_{n/p}$ neutron-to-proton ratio) come with uncorrelated experimental uncertainties $\delta D_i$ propagated from the experimental proton and deuteron data (or $n/d$ and $d/p$ ratio data) and with a procedural uncertainty $\delta^\CJ$ due to data cross-normalization and the treatment of nuclear corrections. 
More specifically, the procedural uncertainty is due to the determination of the normalization factor $\lambda^{\rm norm}$, the systematic nuisance parameters $\lambda_k$, and the use of the deuteron correction factor $R_{d/N}^\CJ(x,Q^2)$. 
These are calculated utilizing the fitted CJ15 PDFs, nuclear corrections, and $1/Q^2$ power corrections, which in turn depend on a number of fitted parameters reflecting the uncertainty of the analyzed experimental data.

The uncertainties in the PDF and other QCD fitting parameters can be propagated into any observable using the Hessian method discussed in Refs.~\cite{Owens:2012bv, Pumplin:2002vw}.
This involves calculating the quantity of interest utilizing a set of ``error PDF sets'', and estimating the uncertainty by comparing the obtained results with the calculation that utilizes the best-fit, or ``central'', PDF set.
For the purpose of the current analysis, we have used a modified version of the published sets of CJ15 error PDFs that include the higher twist correction parameters needed for $F_2$ structure function calculation in addition to the PDFs and off-shell correction parameters, for a total of 24 fitted parameters \cite{Accardi:2016qay}. Each error set was then scaled along its eigendirection to ensure a more faithful determination of the $\Delta \chi^2=1.646$ error band than provided by a straightforward use of the customary Gaussian approximation. This modified PDF set, named \texttt{CJ15nlo\_mod}, along with corresponding calculated DIS structure functions are publicly available, see Appendix~\ref{app:F2grids}.

With the modified PDF set fixed, we have repeated the fit of $\lambda^{\rm norm}$ and $\lambda_k$ and the calculation of $R_{d/N}^\CJ(x,Q^2)$ with each error PDF set in turn, to obtain a set of 49 ($24\times2+1$) values of $n^{(j)}$, $\lambda_k^{(j)}$, and $R_{d/N}^{\CJ,(j)}(x,Q^2)$, with $j=0$ representing the values obtained with the central CJ15 set and $j=1,\ldots,48$ corresponding to the fits obtained with each CJ15 error set.
With these, we can evaluate the uncertainties on any quantities of interest.
For the nuisance parameters $\lambda$, the symmetric CJ uncertainty is defined as 
\begin{equation}
\delta^\CJ \lambda
= \frac{1}{2}
  \sqrt{\sum_{j=1}^{24}
    \Big[ \lambda^{(2j-1)} - \lambda^{(2j)} \Big]^2} \ ,
\label{eq:sym_err_norm}
\end{equation}
and nominally produces 90\% confidence level uncertainties that corresponds to a  $t=1.646$ ``tolerance factor'', see Refs.~\cite{Accardi:2016qay, Accardi:2016ndt} and Appendix~\ref{app:F2grids}. Other confidence levels can be obtained by scaling the result to the desired value of the tolerance.
The resulting $\delta^\CJ \lambda^{\rm norm}$ uncertainties on the normalization nuisance parameters, listed in Tables~\ref{tab:DIS-matched-data} and \ref{tab:DIS-ratio-data}, correspond to normalization uncertainties $\delta^\CJ n$ ranging from 0.1\% to 0.5\%, and are subleading compared to the uncorrelated experimental uncertainties. 
The $\delta^\CJ \Delta_i$ uncertainties on the fitted systematic data shifts are similarly subleading. 

\newpage
The discussed sources of procedural uncertainties can be simultaneously accounted for by repeating the data cross-normalization procedure 48 times with each pair of error PDFs, and defining the procedural uncertainty on the cross-normalized data as
\begin{equation}
\delta^\CJ \widehat D_i
= \frac{1}{2}
  \sqrt{\sum_{j=1}^{24}
    \Big[ \widehat D^{(2j-1)} - \widehat D^{(2j)} \Big]^2} \ .
\end{equation}
This can be (somewhat conservatively) considered to be point-to-point uncorrelated. Finally, the cross-normalized data can be quoted as
\begin{align}
    \widehat D_i = \widehat D_i^{(0)} \pm \delta \widehat D_i \pm \delta^{\CJ} \widehat D_i \ , 
\end{align}
with the uncorrelated uncertainty $\delta \widehat D_i = n^{(0)}\, \delta D_i$ obtained by propagating the experimental value. 
In the results presented here, these two uncertainties are summed in quadrature.

The evaluation of the procedural uncertainty on $F_2^n$ follows in a similar manner, with the neutron extraction repeated 48 times in addition to the determination of its central value, discussed in Sec.~\ref{sec:F2n_extraction},
\begin{equation}
\delta^\CJ \widehat{F}_2^{n}
= \frac{1}{2}
  \sqrt{\sum_{j=1}^{24}
    \Big[ \widehat{F}_2^{n,(2j-1)} -  \widehat{F}_2^{n,(2j)}  \Big]^2} \ .
\label{eq:sym_err}
\end{equation}
As discussed for the data cross-normalization case, this uncertainty is point-to-point uncorrelated, and quoted in addition to the uncorrelated $\delta F_2^n$ experimental errors obtained by straightforward propagation of the uncorrelated $\delta F_2^p$ and $\delta F_2^d$ uncertainties.
We can therefore represent the extracted $F_2^n$ data points and their uncertainties as
\begin{equation}
   \widehat F_2^n=\widehat F_2^{n(0)} \pm \delta \widehat F_2^{n} \pm \delta^\CJ \widehat F_2^{n} \ ,
  \label{eq:F2nhat}
\end{equation}
with the $F_2^n$ results shown in Fig.~\ref{fig:F2n-kin} as a function of $Q^2$ in selected bins of $x$. 
The treatment of the experimental and procedural uncertainties for the neutron-to-proton ratio $R_{n/p}$ (see Fig.~\ref{fig:F2np-kin}) is analogous to that discussed above for the neutron structure function~$F_2^n$. This extracted data is publicly available in the database discussed in Appendix~\ref{app:CJdatabase}.

\subsection{Bin-centered data}
\label{sec:bin_centering}

Bin-centering of the $F_2$ data in $Q^2$ (or $x$) is not only useful graphically, as in Fig.~\ref{fig:f2_bincentered}, but also for applications such as the analysis of the Gottfried sum rule or the evaluation of structure function moments, to be discussed in Sec.~\ref{sec.applications}, which are typically performed at fixed values of $Q^2$.
We perform bin centering in $Q^2$ utilizing the CJ15 structure functions as the underlying model, and defining the bin-centering ratio
\begin{equation}
  R_{\rm bc}(Q_0^2,Q^2)
    \equiv \frac{F_2(x,Q_0^2)}{F_2(x,Q^2)}
    \Bigg|_\CJ \ ,
\end{equation}
where $Q_0$ is the nominal center of the bin, for any given structure function or structure function ratio.
The bin-centered structure functions can then be calculated by multiplying the $\widehat{F}_2$ values by the bin-centering ratio,
\begin{equation}
  \widetilde{F}_2(x,Q_0^2) \equiv R_{\rm bc}(Q_0^2,Q^2)\, \widehat{F}_2(x,Q^2).
\end{equation}
The experimental and procedural uncertainties are propagated from $\widehat{F}_2$ to $\widetilde{F}_2$, for both the proton and neutron structure functions.
The bin-centering procedural uncertainty $\delta_{\rm bc}^\CJ$ is less than 0.1\% and can be safely neglected. Bin centering in $x$ can be performed in an analogous way if needed.

The $Q^2$ bin-centered datasets
\begin{equation}
    \widetilde{F}_2(x,Q_0^2) \pm \delta \widetilde{F}_2 \pm \delta^\CJ \widetilde {F}_2, 
\end{equation}
for the proton, neutron and deuteron structure functions and the $d/p$, $n/d$ and $n/p$ ratios with $Q_0^2=2, 4, 6, 8, 10, 12, 15, 20, 30, 40,$ and 60~GeV$^2$ as in Fig.~\ref{fig:f2_bincentered} are publicly available in the database discussed in Appendix~\ref{app:CJdatabase} alongside their values at the original experimental kinematics.

\newpage
\section{Applications of neutron data}
\label{sec.applications}

In this section we discuss several applications of the extracted neutron dataset to the determination of the isovector nucleon structure function, $F_2^p-F_2^n$, and its lowest moments, including the Gottfried sum rule.

\subsection{Gottfried sum rule}
\label{sec.GSR}

The Gottfried sum is given by the integral over $x$ of the isovector nucleon structure function, $F_2^p-F_2^n$, scaled by the factor $1/x$~\cite{Gottfried:1967kk}.
Since we will be interested also in the shape of the integrand of the Gottfried sum, as well as its saturation as $x \to 0$, it is convenient to define the truncated Gottfried integral
\begin{equation}
\label{eq.IG}
I_G(x_{\rm min},x_{\rm max};Q^2)
= \int_{x_{\rm min}}^{x_{\rm max}} \frac{\dd x}{x}
  \left[ F_2^p(x,Q^2) - F_2^n(x,Q^2) \right],
\end{equation}
so that the Gottfried sum can be expressed as
\begin{equation}
\label{eq.SG}
S_G(Q^2) \equiv I_G(0,1;Q^2).
\end{equation}
At leading order (LO) in the strong coupling $\alpha_s$ and leading power in $1/Q^2$, the $F_2$ structure function can be written in terms of a sum of quark and antiquark PDFs,
    $F_2^{\rm (LO)}(x,Q^2) = x \sum_q e_q^2 (q + \bar{q})(x,Q^2)$.
In this case the Gottfried sum can be written as
\begin{eqnarray}
\label{eq.GSRlo}
S_G^{\rm }(Q^2)
&=& \frac13 \int_0^1 \dd x\, \big( u + \bar u - d - \bar d \big)(x,Q^2)
\notag\\
&=& \frac13 - \frac23 \Delta(Q^2), 
\end{eqnarray}
where the integrated antiquark asymmetry is 
\begin{eqnarray}
\Delta(Q^2) = \int_0^1 \dd x\, \big[ \bar d(x,Q^2) - \bar u(x,Q^2) \big].
\end{eqnarray}
The constant term in the Gottfried integral (\ref{eq.GSRlo}) arises from the normalization of the valence quark distributions,
    $q_v \equiv q - \bar q$, 
equal to 2 or 1 for $q=u$ or $q=d$ quarks in the proton, respectively.
If one further assumes that the $\bar d-\bar u$ difference integrates to zero, one arrives at the canonical (``naive'') Gottfried sum rule prediction of $S_G^{\rm naive} = 1/3$~\cite{Gottfried:1967kk}.

In contrast to many expectations, the value reported by the NMC from an analysis of deep-inelastic muon-hydrogen and muon-deuterium scattering data, \mbox{$S_G^{\rm NMC} = 0.235 \pm 0.026$} \cite{Amaudruz:1991at, Arneodo:1994sh} at $Q^2=4$~GeV$^2$, was significantly lower than the naive flavor symmetric expectation, suggesting a strong violation of the sum rule and the first compelling evidence for $\bar d \not= \bar u$.
This finding prompted tremendous excitement in the nuclear and particle physics community, and spurred considerable work on both the theoretical~\cite{Kumano:1997cy, Geesaman:2018ixo} and experimental~\cite{Baldit:1994jk, Ackerstaff:1998sr, Hawker:1998ty, Towell:2001nh} fronts to better understand this violation.
In particular, since higher order corrections to the Gottfried sum in Eq.~(\ref{eq.GSRlo}) from perturbative gluon radiation were found~\cite{Ross:1978xk} to be very small numerically, it suggested that significant nonperturbative effects were likely to be responsible for the asymmetry.
Such a nonperturbative effect was in fact predicted~\cite{Thomas:1983fh} from chiral symmetry breaking and the associated pion cloud of the nucleon, and has become a standard explanation for the $\bar d$ excess of $\bar u$ in the proton~\cite{Salamu:2014pka, Accardi:2019ofk}. 
A recent analysis also examined the impact of extrapolation methods and higher twist effects on $S_G$~\cite{Kotlorz2021}.

The uncertainty on the NMC value for $S_G$ arises from statistical and systematic errors, including from extrapolations of $F_2^p - F_2^n$ into the unmeasured regions at $x \to 0$ and $x \to 1$. 
The NMC analysis assumed that effects from nuclear corrections to $F_2^n$ extracted from inclusive deuterium data were negligible, defining
    $(F_2^n/F_2^p)_{\rm NMC} \equiv 2F_2^d/F_2^p - 1$,
which assumes that $R_{d/N}=1$.
As discussed in the previous sections, however, nuclear binding and smearing corrections at large $x$~\cite{Melnitchouk:1995fc, Owens:2012bv, Alekhin:2017fpf} and nuclear shadowing at low $x$~\cite{Badelek:1991qa, Zoller:1991ns, Melnitchouk:1992eu, Piller:1995kh} give rise to clear deviations of $R_{d/N}$ from unity.
In the present analysis, considerable effort has been made to account for the nuclear corrections in the extraction of the neutron $F_2$, as discussed in Sec.~\ref{sec.shujie_data_extraction}, and in the following we examine the impact of those corrections on the Gottfried integral.

\begin{figure}[t]
\begin{center}
\includegraphics[width=0.85\textwidth]{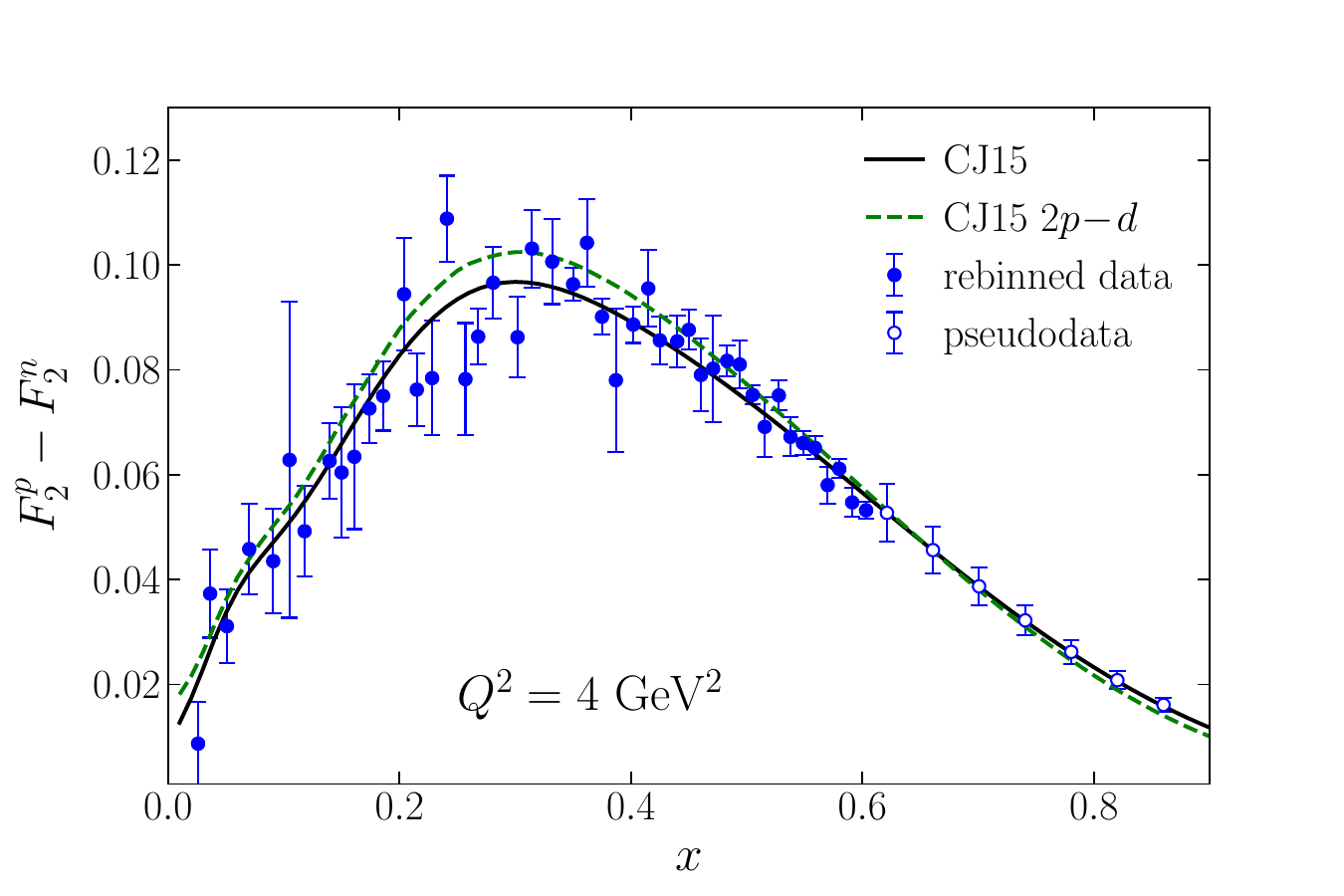}
\end{center}
\vspace*{-0.5cm}
\caption{Isovector nucleon structure function, $F_2^p - F_2^n$, versus $x$ from the combined reanalysis at $Q^2 = 4$~GeV$^2$ (filled circles), compared with the CJ15 global fit (solid black line) and the CJ15 $2F_2^p - F_2^d$ structure function combination (dashed green line), which would coincide with the isovector combination in the absence of nuclear effects. The pseudodata points in the extrapolated region at $x > 0.603$ (open circles) are generated from the CJ15 calculation with 5\% nominal model uncertainties on $F_2^p$ and $F_2^n$. }
\label{fig:F2pn_x}
\end{figure}

The analysis in Sec.~\ref{sec.shujie_data_extraction} combined proton and deuteron $F_2$ structure function measurements from SLAC~\cite{Whitlow1990, Whitlow1992}, BCDMS~\cite{Benvenuti1989, Benvenuti1990}, NMC~\cite{Arneodo1995, Arneodo1997, Arneodo1997a} and JLab~\cite{CLAS:2014jvt} at a set of fixed $Q^2$ values.
An illustration of the isovector nucleon structure function data, shifted to a common value of $Q^2=4$~GeV$^2$, is shown in Fig.~\ref{fig:F2pn_x} for the measured $x$ range between $x=0.009$ and $x=0.603$. 
The plot also estimates the effects of neglecting deuteron corrections in the extraction of the neutron structure function by comparing the isovector $F_2^p - F_2^n$ structure function to the $2F_2^p - F_2^d$ combination of proton and deuteron structure functions.
As one can see, deuteron corrections are generally comparable to the experimental uncertainties in the mid-$x$ region, and tend to suppress the isovector structure function, or equivalently enhance the neutron structure function.

Computing the integral over all $x$ values, as needed for the Gottfried sum [Eq.~(\ref{eq.SG})], requires extrapolating the structure functions beyond the measured region, to $x=0$ and $x=1$.
Although our aim in the combined analysis of the world $F_2^n$ data is to provide the best possible constraints on the structure functions and their moments, with minimal theoretical bias, such extrapolations will inevitably introduce model dependence into the procedure, especially for the $x \to 0$ behavior.
We will discuss the uncertainties introduced into the extracted functions and moments from both the $x \to 0$ and $x \to 1$ extrapolations in the following.

\begin{figure}[t]
\begin{center}
\includegraphics[width=0.85\textwidth]{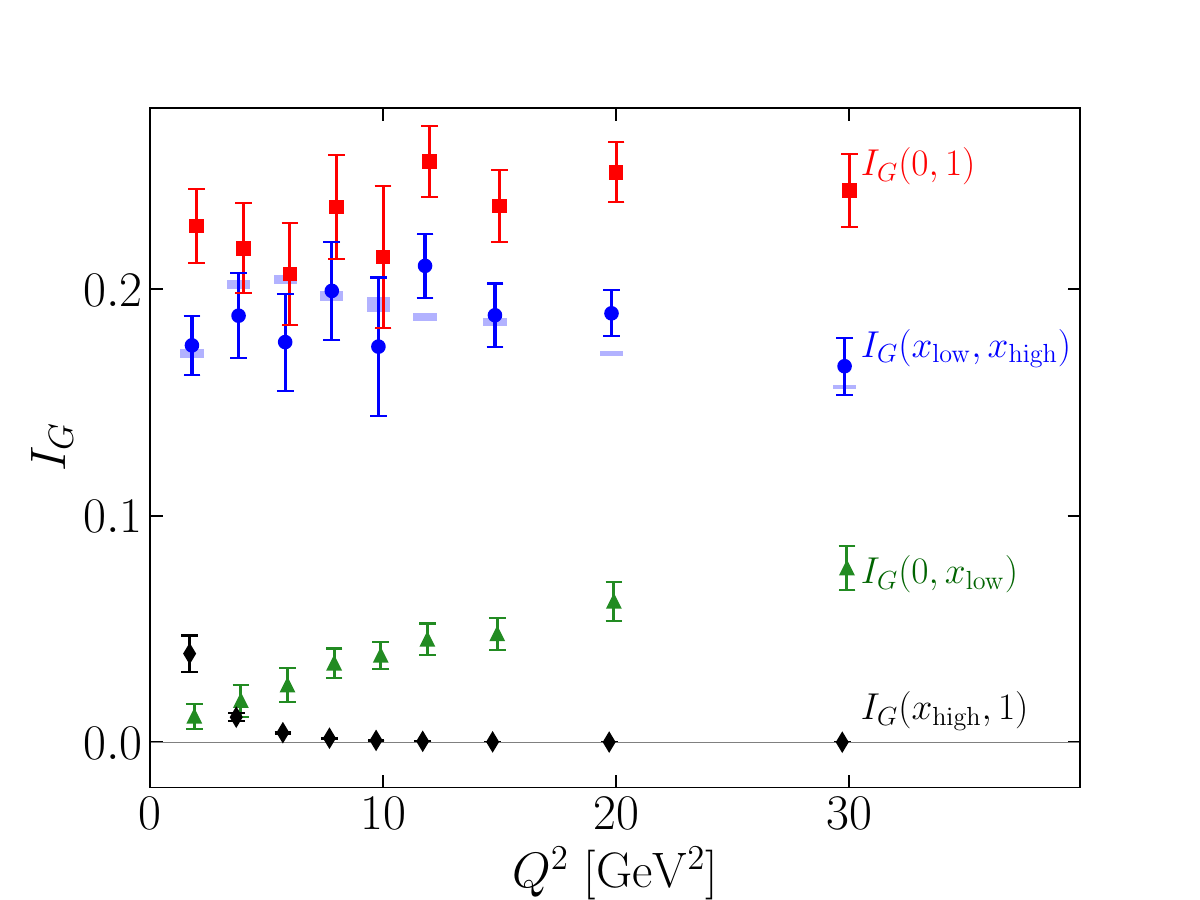}
\end{center}
\vspace*{-0.5cm}
\caption{Variation with $Q^2$ of the contributions to the Gottfried integral $I_G(x_{\rm min},x_{\rm max};Q^2)$ from different $x$ intervals, including from the measured (blue circles), $x \to 0$ extrapolated (green triangles), and $x \to 1$ extrapolated (black diamonds) regions, along with the total integral (red squares). The blue shaded bands show the integral in the measured region calculated from the CJ15 PDFs.}
\label{fig:q2_dep}
\end{figure}

\begin{table}[b]
\centering
\caption{\setstretch{1.0} 
Contributions $I_G(x_{\rm min},x_{\rm max};Q^2)$ to the Gottfried integral from different intervals of $x$ at fixed $Q^2$ values, along with the total integral, $S_G(Q^2)$. The experimentally measured region corresponds to $(x_{\rm low}, x_{\rm high})$. The range $(0.004,1)$ is provided for comparison with the NMC experiment~\cite{Amaudruz:1991at, Arneodo:1994sh}.\\}
\scalebox{0.9}{
\begin{tabular}{cccccccc} \hline
&  &  & \multicolumn{4}{c}{$I_G$} & \\ \cline{4-7} 
$Q^2$ (GeV$^2$)~~ & ~$x_{\rm low}$~ & ~$x_{\rm high}$~ &
$(0,x_{\rm low})$~ &
$(x_{\rm low}, x_{\rm high})$~ &
~$(x_{\rm high}, 1)$~ &
~$(0.004, 1)$ &
$S_G$  \\  \hline \\
2	&0.005	&0.422	&0.011(6)	&0.175(13)	&0.039(8)	&0.218(16) &\textbf{	0.228(16)	}\\ \\
4	&0.009	&0.603	&0.018(7)	&0.188(19)	&0.011(2)	&0.207(19) &\textbf{	0.218(20)	}\\ \\
6	&0.014	&0.690	&0.025(8)	&0.177(21)	&0.004(0)	&0.195(22) &\textbf{	0.207(23)	}\\ \\
8	&0.024	&0.747	&0.035(7)	&0.199(22)	&0.002(0)	&0.224(22) &\textbf{	0.236(23)	}\\ \\
10	&0.028	&0.781	&0.038(6)	&0.175(31)	&0.001(0)	&0.204(33) &\textbf{	0.214(31)	}\\ \\
12	&0.035	&0.819	&0.045(7)	&0.210(14)	&0.000(0)	&0.245(15) &\textbf{	0.256(16)	}\\ \\
15	&0.037	&0.851	&0.048(7)	&0.189(14)	&0.000(0)	&0.225(14) &\textbf{	0.237(16)	}\\ \\
20	&0.053	&0.877	&0.062(9)	&0.189(10)	&0.000(0)	&0.240(11) &\textbf{	0.252(13)	}\\ \\
30	&0.072	&0.896	&0.077(10)	&0.166(13)	&0.000(0)	&0.232(14) &\textbf{	0.243(16)	}\\ \\
\hline
\end{tabular}}
\label{tab:tableIG}
\end{table}

To proceed, we consider the contributions to the total Gottfried integral (\ref{eq.SG}) from individual $x$ regions at a fixed value of $Q^2$,
\begin{equation}
\label{eq:gsr_def3}
S_G(Q^2) 
= I_G(0, x_{\rm low};Q^2)
+ I_G(x_{\rm low}, x_{\rm high};Q^2)
+ I_G(x_{\rm high}, 1; Q^2),
\end{equation}
where $x_{\rm low}$ and $x_{\rm high}$ are the lower and upper bounds of the experimental data, both of which vary with $Q^2$ (see Table~\ref{tab:tableIG}).
The smallest value of $x_{\rm low}$ for any of the bins from the NMC data~\cite{Arneodo1995, Arneodo1997, Arneodo1997a} is $x = 0.004$.
The individual contributions to $S_G(Q^2)$, along with the total extrapolated moments, are displayed in Fig.~\ref{fig:q2_dep} for several $Q^2$ values between $Q^2=2$ and 30~GeV$^2$. 
From kinematics one finds that the values of both $x_{\rm low}$ and $x_{\rm high}$ increase with increasing $Q^2$.
This has the effect of increasing the relative contribution from the extrapolated region at low $x$, and decreasing the relative contribution from the large-$x$ extrapolation.

For the measured region $I_G(x_{\rm low}, x_{\rm high}; Q^2)$, the uncertainties were estimated using $\gtrsim 10^4$ Monte Carlo simulations based on the total systematic and statistical errors reported by each experiment, as discussed in Sec.~\ref{sec.shujie_data_extraction}.
At the $i$-th iteration, each data point $D_j$ was shifted by $\lambda_i \delta^{\,\rm sys}_j$, where $\lambda_i$ is a random number generated by the standard normal distribution, and $\delta^{\,\rm sys}_j$ is the total systematic uncertainty of the $j$-th data point. 
All points with their statistical uncertainties were then binned in $x$ with a bin size {$\Delta x = 10^{-4}$}.
A trapezoid integration method with adaptive step size was used with the rebinned data points to obtain an integral $I_i$ $(i=1,\ldots,10^4)$.
The mean and standard deviation of $I_i$ were recorded as the central value and total error of $I_G(x_{\rm low}, x_{\rm high}; Q^2)$, respectively.
In practice, the step size $\Delta x$ was varied between $10^{-5}$ and $10^{-1}$ to check the integral stability, with $\Delta x=10^{-4}$ found to be stable and in good agreement with corresponding analytical results.

For the unmeasured low-$x$ region contribution, $I_G(0, x_{\rm low}; Q^2)$, we followed previous analyses~\cite{Arneodo1995, Arneodo1997, Arneodo1997a, Abbate:2005ct} by assuming a parametrization for the nonsinglet structure function inspired by Regge theory, with functional form
    $F_2^p - F_2^n = A x^\alpha$.
The parameters $A$ and~$\alpha$ were fitted to data at $0.01 < x < 0.1$ for each value of $Q^2$.
We found good fits to the data with $\alpha = 0.6$, which also described well the nonsinglet structure function calculated perturbatively with the CJ15 PDFs in the unmeasured $x < x_{\rm min}$ region. 
The corresponding contributions to GSR integrals were then calculated analytically.

At the lowest $Q^2$ value, $Q^2=2$~GeV$^2$, the unmeasured region contributes $\approx 5\%$ of the total integral.
This fraction becomes larger with increasing $Q^2$ due to the more restricted range of low-$x$ data accessible at the higher $Q^2$ values (see Fig.~\ref{fig:q2_dep}), rising up to $\approx 13\%$ at $Q^2=10$~GeV$^2$.
Since the $x \to 0$ extrapolated contribution to the Gottfried integral is not negligible, it is important to estimate the systematic uncertainty arising from this component.
For this purpose, the value of $\alpha$ was varied between 0.5 and 0.7, and the resulting differences in the truncated GSR integral taken as a measure of the systematic uncertainty.
A small contribution from the uncertainty in the normalization parameter $A$ was also folded into the total error.
For all $Q^2$ points considered, the magnitude of the uncertainty was found to be $\approx 3\%-4\%$ of the total integral.

\begin{figure}[t]
\begin{center}
\includegraphics[width=0.8\textwidth]{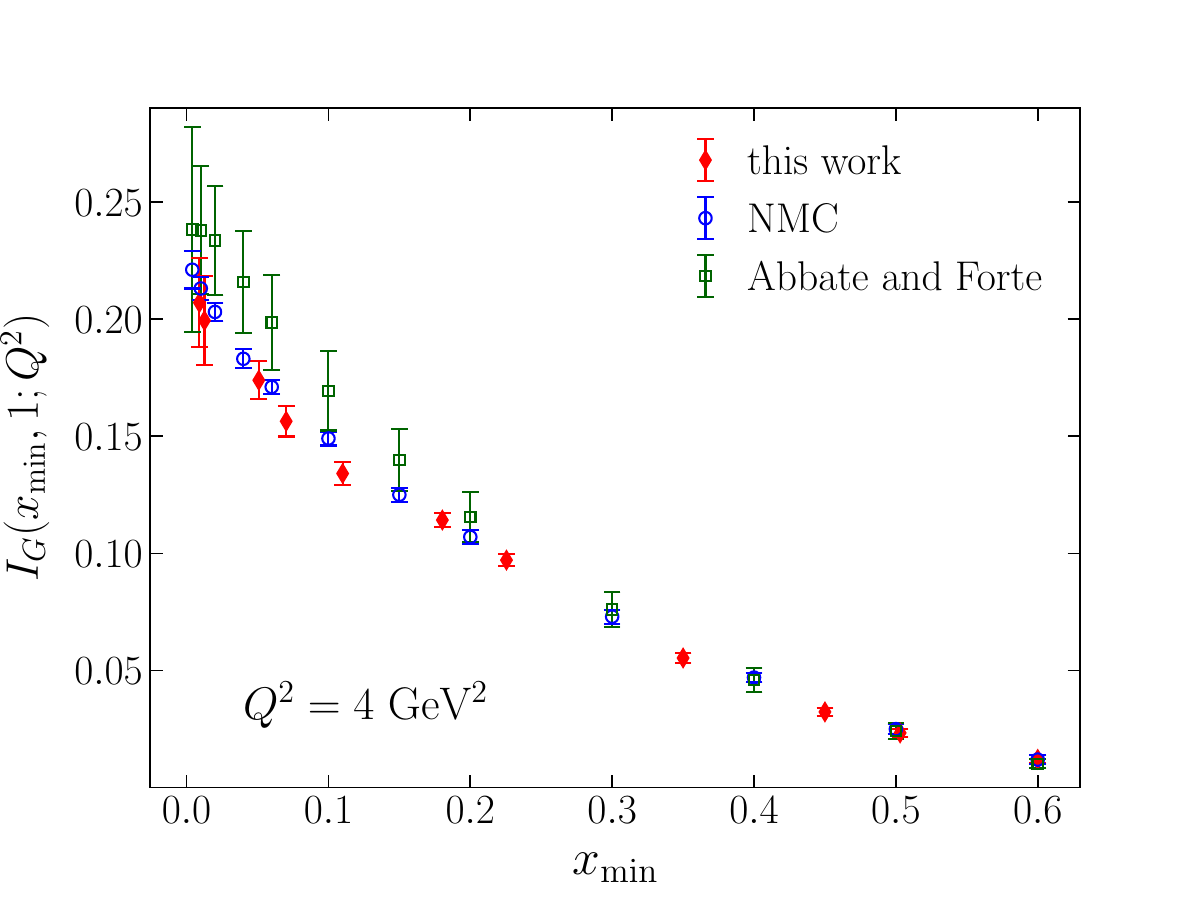}
\end{center}
\vspace*{-0.5cm}
\caption{Dependence of the Gottfried integral, $I_G(x_{\rm min},1;Q^2)$, on the lower limit of the $x$ integration, $x_{\rm min}$, at $Q^2=4$~GeV$^2$ in our analysis (red diamonds), compared with the NMC extraction~\cite{Arneodo1997, Arneodo1997a} (open blue circles) and the analysis by Abbate and Forte~\cite{Abbate:2005ct} (open green squares).}
\label{fig:low_lim}
\end{figure}

The importance of the low-$x$ contribution to the Gottfried integral is also evident in Fig.~\ref{fig:low_lim}, which shows the dependence of $I_G(x_{\rm min},1;Q^2)$ on the lower limit, $x_{\rm min}$, of the $x$ integration at a fixed value of $Q^2=4$~GeV$^2$.
The results are compared with the original extraction from the NMC data analysis~\cite{Arneodo1997, Arneodo1997a}, as well as from the more recent analysis of Abbate and Forte (AF)~\cite{Abbate:2005ct}, and illustrate the saturation of the integral as $x_{\rm min} \to 0$.
While the results from our global reanalysis agree with the previous results at higher $x_{\rm min}$ values, $x_{\rm min} \gtrsim 0.2$, at lower~$x_{\rm min}$ our extracted integrals are slightly below both the earlier NMC and AF results. 
In our analysis we also removed shadowing corrections from the deuterium data, resulting in a 1\% -- 2\% increase of $F_2^d$ at $x \lesssim 0.1$, and hence a comparable increase in the extracted $F_2^n$ and a decrease of the isovector $F_2^p - F_2^n$ and the corresponding integrated value in the same region of $x$.

For the unmeasured large-$x$ region, $x_{\rm high} < x < 1$, the inelastic contribution to the integral was evaluated from the CJ15 PDFs~\cite{Accardi:2016qay}, as well as an empirical fit that describes structure function data in both the DIS and resonance regions~\cite{Christy_2010, Bosted_2008}.
The latter naturally cuts off at $x$ values larger than the pion threshold, $x_\pi = Q^2/(W_\pi^2 - M^2 + Q^2)$, where $W_\pi = M + m_\pi$ is the minimum mass of the inelastic final state.
The former suffers from the ``threshold problem'' affecting TMC calculations in momentum space \cite{Schienbein:2007gr, Accardi:2008ne, Steffens:2012jx, Blumlein:2012bf}, and the integrals are limited to $x < x_\pi$.
In either case, the elastic contribution to the Gottfried integral was not included.
The average of the two calculated values gives the central value of the high-$x$ contribution, and their difference provides an estimate of the systematic extrapolation uncertainty.

\begin{figure}
\begin{center}
\includegraphics[width=0.8\textwidth]{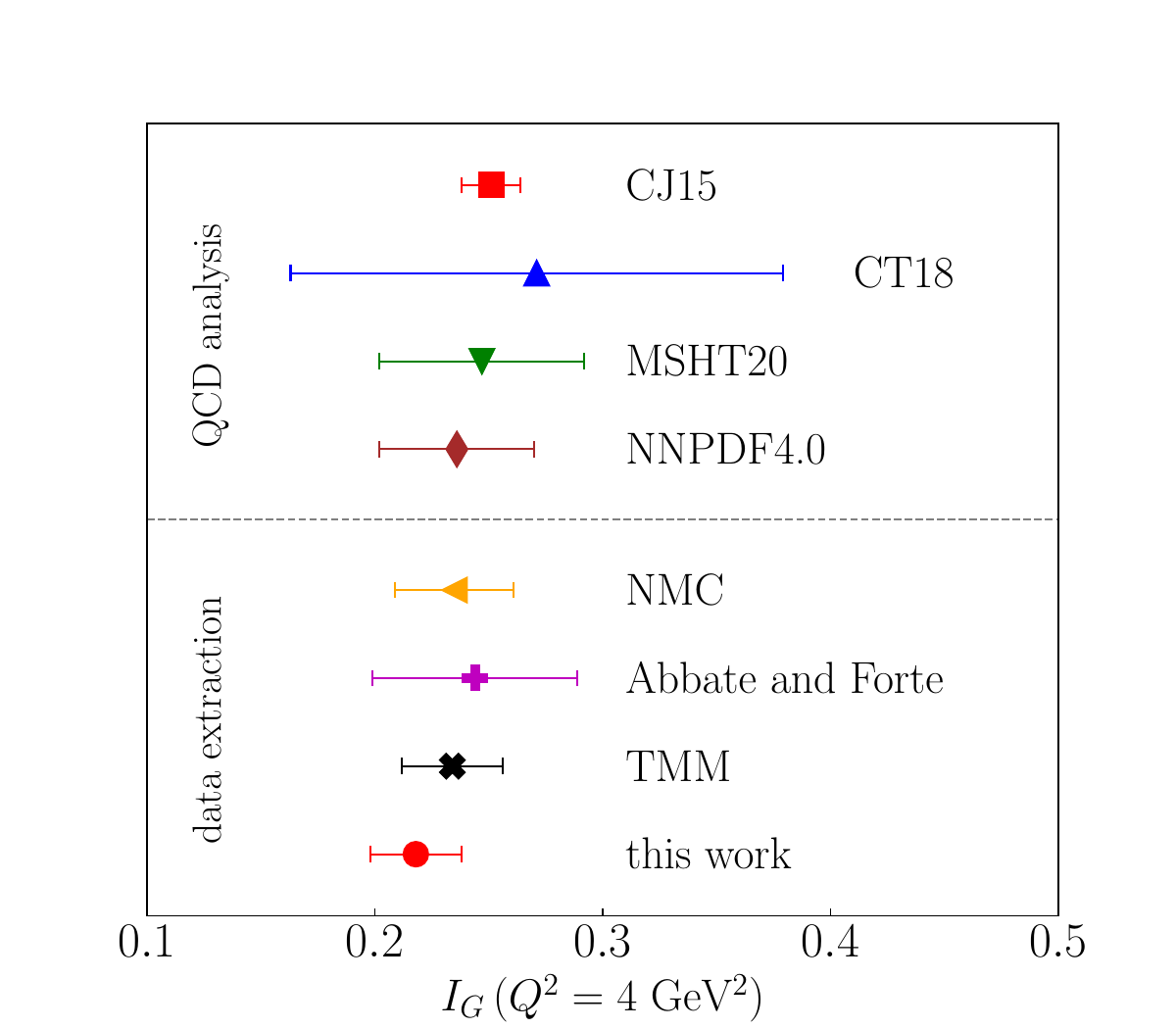}
\end{center}
\vspace*{-0.5cm}
\caption{Gottfried integral $I_G(0,1;Q^2=4$~GeV$^2)$ from data extractions (bottom), and from theoretical calculations using PDF obtained in recent global QCD analyses (top). The smaller CJ15 uncertainty compared to the other QCD calculations is due to the collaboration's choice of a nominal $t=1.646$ tolerance factor, which smaller than in the dynamical tolerance criteria adopted in the other analyses. The large CT18 uncertainty is due to bigger PDF uncertainties at $x\lesssim 0.001$ compared to NNPDF4.0 and MSHT20.}
\label{fig:gsr_compare}
\end{figure}

The final results for the contributions to the Gottfried integral from the various regions of $x$ are summarized in Table~\ref{tab:tableIG}.
At a reference scale of $Q^2=4$~GeV$^2$, we find the contribution from the region $x > 0.004$, corresponding to the range given in the NMC analysis~\cite{Arneodo:1994sh}, to be
    $I_G(0.004,1) = 0.207(19)$, 
where for notational convenience we suppress the $Q^2$ dependence in $I_G$.
This is slightly smaller than the value found in the original NMC data analysis~\cite{Arneodo1995},
    $I_G^{\rm NMC}(0.004,0.8) = 0.221(21)$,
but is consistent within the uncertainties.
It is also compatible with the more recent AF determination~\cite{Abbate:2005ct},
    $I_G^{\rm AF}(0.004,0.8) = 0.228(44)$,
which has a somewhat larger uncertainty.
Including the contributions from the extrapolation into the unmeasured regions $x \to 0$ and $x \to 1$,
we find the total Gottfried integral at $Q^2=4$~GeV$^2$ to be
    $S_G = 0.218(20)$,
which again is slightly smaller than the values extracted from the NMC data alone,
    $S_G^{\rm NMC} = 0.235(26)$
\cite{Arneodo1995}, or in the AF analysis,
    $S_G^{\rm AF} = 0.242(48)$
\cite{Abbate:2005ct}, or from the recent reanalysis of NMC data with the truncated methods,
    $S_G^{\rm TMM} = 0.234(22)$~\cite{Kotlorz2021}.
A comparison of these data extractions is shown in Fig.~\ref{fig:gsr_compare}, along with calculations using global QCD analyses from CJ15~\cite{Accardi:2016qay}, CT18~\cite{Hou:2019efy}, MSHT20~\cite{Bailey:2020ooq}, and NNPDF4.0~\cite{Butterworth:2015oua}.\footnote{The calculations were done with open-source code \textsc{APFEL++}~\cite{Bertone:2013vaa,Bertone:2017gds} using PDF sets from the \textsc{LHAPDF} library~\cite{Buckley:2014ana,Andersen:2014efa}.}
Taken collectively, these latter values appear to slightly overestimate the data extractions.

To estimate the higher twist contributions to the Gottfried integral, we compare the results using the CJ15 calculation including and neglecting the subleading $1/Q^2$ corrections to the structure functions.
We find effects for $S_G$ of the order 0.1\%, so that to a good approximation we can use Eq.~(\ref{eq.GSRlo}) to convert our extracted $S_G$ value to an integrated LO flavor asymmetry in the proton of
    $\Delta = 0.173(30)$ 
at $Q^2=4$~GeV$^2$, which is slightly larger than the NMC result
    $\Delta^{\rm NMC} = 0.148(39)$
\cite{Arneodo1995}, but consistent within the quoted uncertainties.
It is also somewhat larger than the integrated asymmetry extracted from the E866 Drell-Yan data~\cite{Towell:2001nh} (which were used to constrain the $\bar d/\bar u$ ratio) when combined with the parametrization of $\bar d + \bar u$ from the CTEQ5 global QCD analysis~\cite{Lai:1999wy},
    $\Delta^{\rm E866} = 0.118(12)$.
Our reanalysis of the neutron structure function data therefore suggests a stronger violation of SU(2) flavor asymmetry in the proton sea than reported in the previous studies.

Within the chiral effective field theory framework of Refs.~\cite{Salamu:2014pka, Salamu:2018cny}, in which the nonperturbative sea in the proton is generated from pseudoscalar pion loops, an integrated asymmetry of $\Delta \approx 0.18$ would correspond to an average $\pi^+$ multiplicity in the quantum fluctuation of a proton to a $\pi^+$ and neutron state of $\langle n \rangle_{\pi^+ n} \approx 0.25$.
In comparison, a smaller value of $\Delta \approx 0.12$ as obtained from the E866 analysis~\cite{Towell:2001nh} would correspond to a $\pi^+$ multiplicity of $\langle n \rangle_{\pi^+ n} \approx 0.15$.
The larger deviation from the Gottfried sum rule observed in our analysis implies therefore a $\gtrsim 50\%$ larger pion cloud than that suggested by the previous studies.

\subsection{Nonsinglet moments}

In recent years developments in lattice QCD have enabled precision calculations of moments of PDFs from first principles~\cite{Lin:2017snn, Constantinou:2020hdm}.
Comparison of the calculated nonsinglet quark distribution moments can be made with moments extracted from experimental data, providing a valuable test of the lattice methodology and various high order corrections, as well as of QCD itself.
The nonsinglet moment of the $u$ and $d$ quark PDFs accessible to lattice QCD is given by
\begin{equation}
    \langle x \rangle_{u^+-d^+} = \int \dd x \, x\,
    \big[ u(x) + \bar{u}(x) - d(x) - \bar{d}(x) \big],
\label{eq:nonsinglet-PDF-moment}
\end{equation}
which corresponds to the difference between the momentum carried by $u$ and $d$ quarks in the proton, computed at some resolution scale usually set by the lattice spacing.
For a direct comparison between the calculated PDF moments and those extracted from experiment, we consider the Nachtmann moment of the $F_2^N$ structure function of the nucleon $N$, which accounts for kinematical target mass effects associated with higher spin operators,
\begin{equation}
    M_2^{p-n}(Q^{2})
    = \int_0^1 \dd x\, \frac{\xi^3}{x^3 } 
    \bigg[ \frac{3 + 9r + 8r^2}{20} \bigg]
    F_2^{p-n}(x,Q^2),
    \label{eq:F2_Nachtmann_moment}
\end{equation}
where
$\xi = 2x/(1 + r)$ is the Nachtmann scaling variable~\cite{Nachtmann:1973mr, Greenberg:1971lpf}, with $r = \sqrt{1 + 4 M^2 x^2/Q^2}$.
We neglect the small difference between the proton and neutron masses, by setting the nucleon mass to $M=0.939$~GeV.

\begin{figure}
\begin{center}
\includegraphics[width=0.8\textwidth]{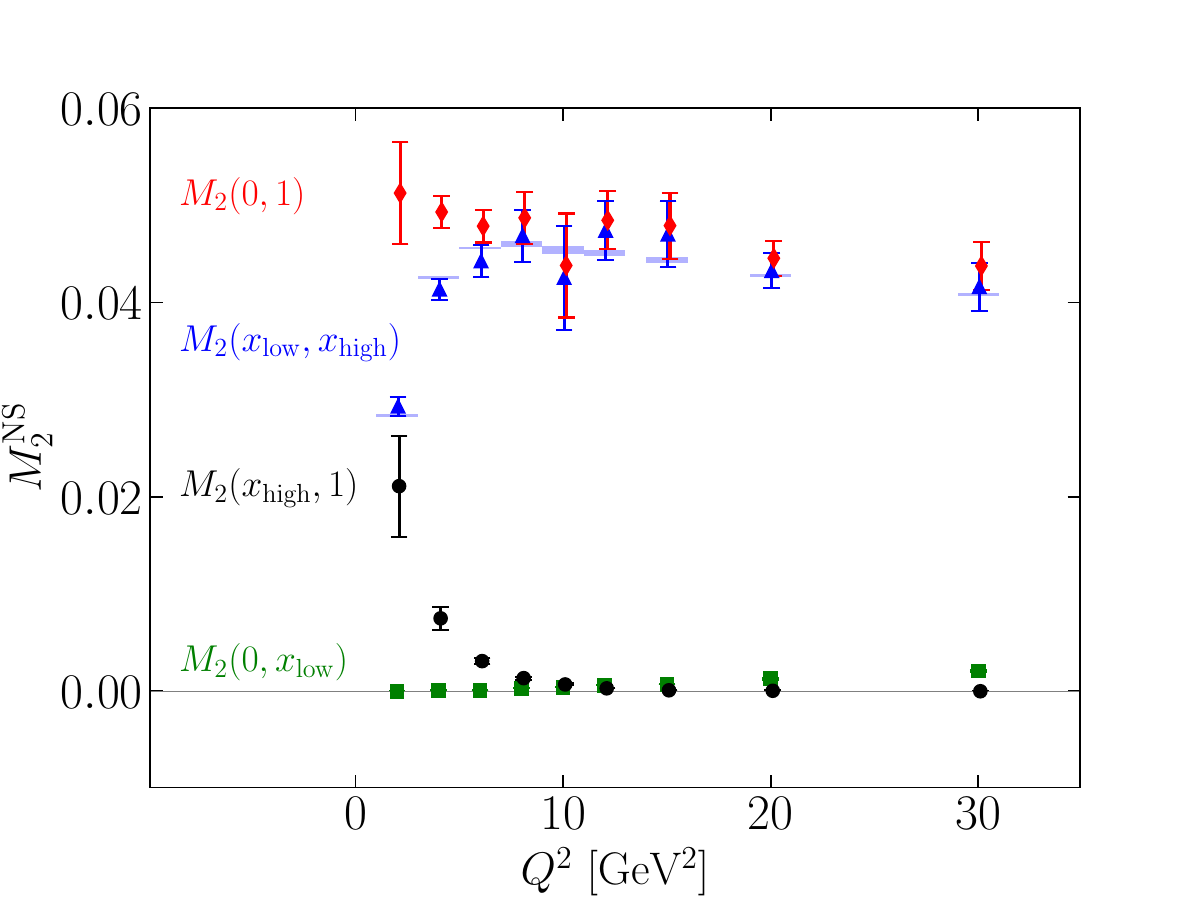}
\end{center}
\vspace*{-0cm}\caption{\setstretch{1.0} Nonsinglet Nachtmann moment $M_2^{\rm NS}$ extracted from world DIS data at various values of $Q^2$. The total moments (red diamonds) are broken down into contributions from different $x$ intervals: the measured region $[x_\text{low},x_\text{high}]$ (blue triangles), the $x > x_\text{high}$ extrapolation region (black circles), and the $x < x_\text{low}$ extrapolation region (green squares). The blue shaded bands represent the integral in the measured region calculated using CJ15 structure functions instead of data points.}
\label{fig:momenta-Q2}
\end{figure}

Using the same rebinned $F_2^p$ and $F_2^n$ datasets used in the GSR analysis and deploying the integration technique described in the previous section to the integral in Eq.~\eqref{eq.IG}, we then extract the nonsinglet moments,
\begin{align}
    M_2^{\rm NS}(Q^2)= M_2^p(Q^2) - M_2^n(Q^2),
\end{align}
from the world DIS data. 
To gauge the relative contribution of the measured and extrapolation regions to the full moment, we also compute the truncated moments $M_2^{\rm NS}(x_{\rm min},x_{\rm max};Q^2)$ by restricting the $x$ integration in Eq.~\eqref{eq:F2_Nachtmann_moment} to the $[x_{\rm min},x_{\rm max}]$ interval.
The results, including the contributions from each of the integration regions, are shown in Fig.~\ref{fig:momenta-Q2} as a function of $Q^2$ and listed in Table~\ref{tab:tableM2}. 
The low-$x$ extrapolation region gives a negligible contribution to the moments, while the contribution from the large-$x$ region increases with decreasing $Q^2$; indeed, the unmeasured large-$x$ interval widens as $Q^2$ becomes smaller.
As with the GSR, the extracted moments are only weakly dependent on $Q^2$, suggesting a possibly large cancellation between the proton and neutron HT components.

\begin{table}
\centering
\caption{\setstretch{1.0} 
Contributions to the nonsinglet Nachtmann moment $M_2^{\rm NS}(x_{\rm min},x_{\rm max};Q^2)$ from various intervals of $x$ at fixed $Q^2$ values from 2 to 10~GeV$^2$, along with the total integral. The experimentally measured region corresponds to $(x_{\rm low}, x_{\rm high})$. Analysis methods are the same as described in Sec.~\ref{sec.GSR}. The column labelled ``HT'' gives the higher twist contribution to the total integral $M_2^{\rm NS}(0,1;Q^2)$ as a percentage.\\}
\scalebox{0.9}{
\begin{tabular}{cccccccc} \hline
&  &  & \multicolumn{4}{c}{$M_2$} & \\ \cline{4-7} 
$Q^2$ (GeV$^2$)~~ & ~$x_{\rm low}$~ & ~$x_{\rm high}$~ &
$(0,x_{\rm low})$~ &
$(x_{\rm low}, x_{\rm high})$~ &
~$(x_{\rm high}, 1)$~ &
$(0,1)$ &
~HT (\%)  \\  \hline \\
2	&	~0.005~	&	~0.408~	&	0.0000(0)	&	0.0293(10)	&	~0.0211(52)	&	\textbf{~0.0513(53)}	&	3.7	\\\\
4	&	~0.009~	&	~0.603~	&	0.0001(0)	&	0.0414(11)	&	~0.0075(12)	&	\textbf{~0.0493(16)}	&	1.1	\\ \\
6	&	~0.014~	&	~0.690~	&	0.0001(0)	&	0.0442(16)	&	0.0031(3)	&	\textbf{~0.0479(17)}	&	0.5	\\\\
8	&	~0.024~	&	~0.747~	&	0.0003(0)	&	0.0467(26)	&	0.0014(2)	&	\textbf{~0.0487(27)}	&	0.3	\\\\
10	&	~0.028~	&	~0.791~	&	0.0004(0)	&	0.0425(52)	&	0.0007(1)	&	\textbf{~0.0438(54)}	&	0.2	\\ \\
12  &   ~0.035~ &   ~0.819~ &   0.0006(0)	& 0.0474(30)    &    0.0003(0)    &   \textbf{~0.0485(30)}    &   0.2\\ \\
15  &   ~0.037~ &   ~0.851~ &   0.0007(0)	& 0.0470(34)    &    0.0001(0)    &   \textbf{~0.0479(34)}    &   0.2\\ \\
20  &   ~0.053~ &   ~0.877~ &   0.0013(1)	& 0.0433(18)    &    0.0001(1)    &   \textbf{~0.0446(18)}    &   0.1\\ \\
30  &   ~0.072~ &   ~0.896~ &   0.0021(1)	& 0.0416(25)    &    0.0000(0)    &   \textbf{~0.0438(25)}    &   0.1\\ \\\hline
\end{tabular}}
\label{tab:tableM2}
\end{table}

Within the QCD factorization approach, the nonsinglet Nachtmann moment $M_2^{\rm NS}$ can be related to the nonsinglet moment $\langle x \rangle_{u^+-d^+}$ of the quark PDFs by dividing out the perturbative Wilson coefficient, and subtracting possible HT contributions,
\begin{align}
    \frac{3}{C_2} M_2^{\rm NS} & = \langle x \rangle_{u^+-d^+} + \text{HT} \ ,   
\label{eq:moment-to-PDF}
\end{align}
where to ${\cal O}(\alpha_s)$ the Wilson coefficient $C_2 = 1 + 1.0104 \, \alpha_s(Q^2) / 4\pi$ \cite{Weigl:1995hx}.
We can estimate the size of HT contribution by calculating the CJ15-calculated $M_2^{\rm NS}$ with and without HT corrections, reporting the relative effect in Table~\ref{tab:tableM2}. 
We find that the HT contribution is typically smaller than the extracted data uncertainty, and becomes comparable to this only at the lowest $Q^2$ value.

\begin{figure}
    \begin{center}
    \includegraphics[width=0.85\textwidth]{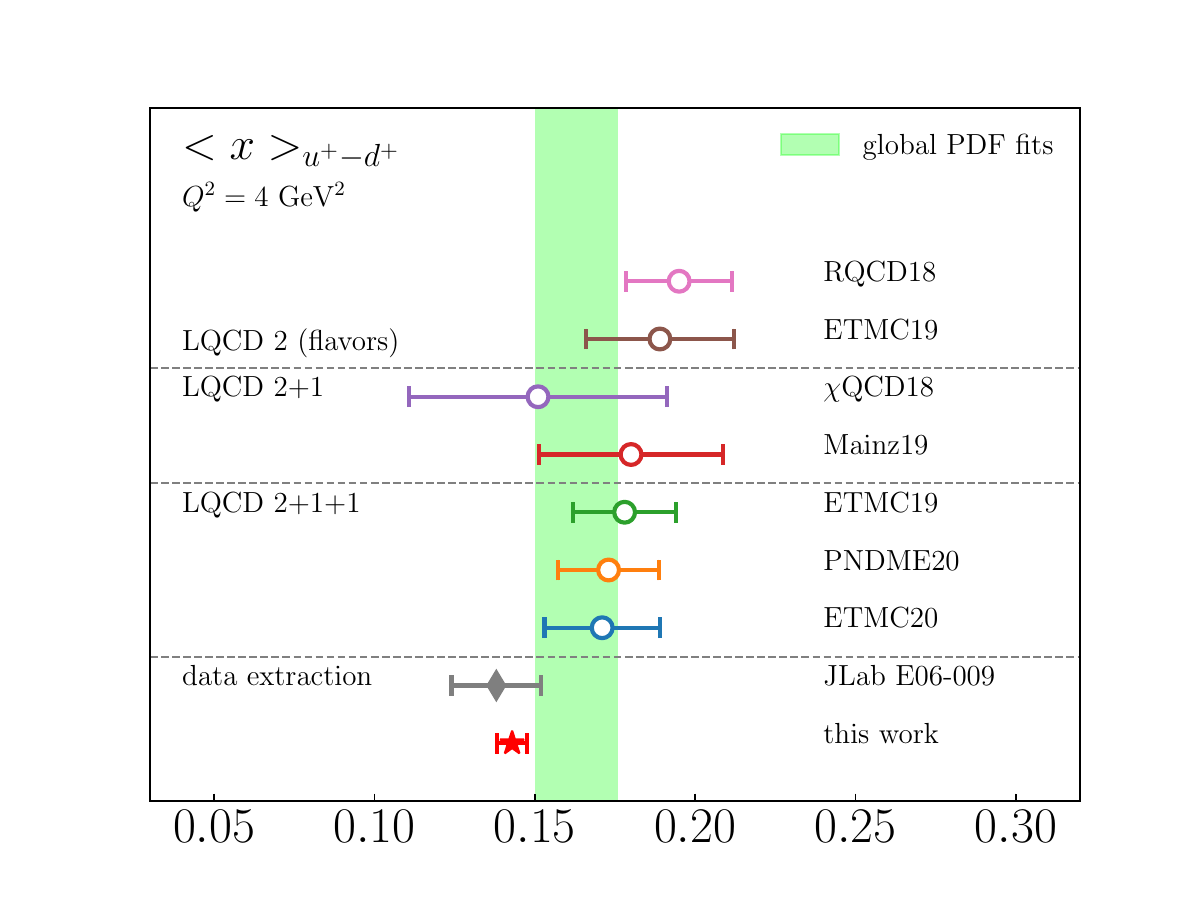}
    \end{center}
    \vspace*{-1cm}\caption{\setstretch{1.0} A comparison of the nonsinglet PDF moment $\langle x\rangle_{u^+-d^+}$ extracted from DIS structure functions data (solid symbols) and from various lattice QCD calculations (open circles) at $Q^2=4$~GeV$^2$. The green vertical band shows the unweighted combination of moments from recent global PDF analyses discussed in the text. }
\label{fig:momenta-stacked}
\end{figure}

In particular, at $Q^2=4$~GeV$^2$ the HT contribution is negligible compared to the $M_2^{\rm NS}$ extraction uncertainties, and inverting Eq.~\eqref{eq:moment-to-PDF} we find $\langle x \rangle_{u^+-d^+} = 0.143(5)$ at this scale.
This result is shown in Fig.~\ref{fig:momenta-stacked} compared with a recent experimental extraction from precision proton and deuteron $F_2$ structure function data from the E06-009 experiment at JLab~\cite{Albayrak2018}, and with various lattice QCD calculations as reviewed in Refs.~\cite{Lin:2017snn, Constantinou:2020hdm}.
In addition, our extracted moment is also compared with an average of PDF nonsinglet moments using PDFs parametrizations from recent global QCD analyses, using Eq.~\eqref{eq:nonsinglet-PDF-moment} directly.
The corresponding numerical values of all the moments are given in Table~\ref{tab:moments}.

\begin{table}[b]
\centering
\caption{\setstretch{1.0} Nonsinglet $\langle x \rangle_{u^+-d^+}$ PDF moments at $Q^2=4$~GeV$^2$ obtained from DIS data extractions, lattice QCD calculations, and PDF global analyses. The lattice values are envelopes of the calculations shown in Fig.~\ref{fig:momenta-stacked} for each quark flavor scheme, and the global QCD analyses results are described in the main text.\\}
\begin{tabular}{ccccccc} 
\hline\hline
      &  \multicolumn{2}{c}{\bf data extractions}  
      && \multicolumn{3}{c}{\bf lattice QCD envelope}
\\ \cline{2-3} \cline{5-7}
      & \small this work &  \small E06-009  
      && \small $N_f$=2 & \small $N_f$=2+1 & \small $N_f$=2+1+1  
\\ 
      &  ~0.143(5)~  & ~0.138(14)~ &&  ~0.189(23)~   & ~~0.160(49)~ & ~0.173(21)~
\\ \hline\hline
\end{tabular}

\begin{tabular}{cccccccc}
\\ \hline\hline
      \multicolumn{8}{c}{\bf global PDF analyses}  
\\ \hline
      \small MMHT2014 & \small CJ15 &\small NNPDF3.1 & \small CT18 & \small JAM21 & \small ABMP2016 & \small HERAPDF2.0 & \small Combined  
\\ 
      0.151(4) & 0.152(2) & 0.152(3) & 0.156(7) & ~0.157(2) & 0.167(4) & 0.188(3) & 0.156$^{(+20)}_{(-6)}$
\\ \hline\hline
\end{tabular}
\label{tab:moments}
\end{table}

In the analysis of the E06-009 data~\cite{Albayrak2018}, nuclear corrections were removed from the deuteron data by adopting the same nuclear convolution model as that utilized here, but using as input the phenomenological structure functions that were fitted directly to inclusive DIS data~\cite{Christy_2010}.
The E06-009 analysis also considered the elastic contribution to the nonsinglet moment and resulted in a value of 
    $\langle x \rangle_{u^+-d^+}^{\rm \mbox{\tiny E06-009}} = 0.138(14)$.
With the elastic contribution removed, this would be reduced by 3\% to 
    $\langle x \rangle_{u^+-d^+}^{\rm \mbox{\tiny E06-009}} = 0.133(14)$, 
which is well within the quoted uncertainties.
As can be appreciated from Fig.~\ref{fig:momenta-stacked}, the E06-009 moment is consistent with that found in our analysis, albeit with a larger uncertainty.

The nonsinglet PDF moments extracted from experimental data appear systematically below those from lattice QCD calculations.
This is clear from the comparisons in Fig.~\ref{fig:momenta-stacked}, where recent lattice QCD moments \cite{RQCD18, xQCD18, Mainz19, ETMC19, ETMC20, PNDME20} have been extrapolated to the physical pion mass,
as reviewed in Ref.~\cite{Constantinou:2020hdm}.
The small tension with the data extraction may be indicative of residual unaccounted for systematic effects in the lattice extraction of the PDF moments.

The nonsinglet moments $\langle x \rangle_{u^+-d^+}$ obtained from an average of recent global QCD analyses, namely, the ABMP16 \cite{Alekhin:2017kpj}, CJ15 \cite{Accardi:2016qay}, CT18 \cite{Hou:2019efy}, HERAPDF2.0 \cite{Abramowicz:2015mha}, JAM19 \cite{Sato:2019yez}, MMHT2014 \cite{Harland-Lang:2014zoa}, and NNPDF3.1 \cite{NNPDF:2017mvq} PDF parametrizations, are compared in Fig.~\ref{fig:momenta-stacked} with some recent lattice QCD simulations~\cite{Lin:2017snn, Constantinou:2020hdm}, and extractions from data.
How to average observables calculated from different global PDF fits is an open question~\cite{Butterworth:2015oua, Accardi:2016ndt}, and in this analysis we quote the median of the individual central values as the central value of the combined moment, and asymmetric standard deviations from the median as systematic errors.
The statistical uncertainties on the individual calculations do not vary substantially, and a simple average represents these well. 
The averaged result is found to be 
    $\langle x \rangle_{u^+-d^+}^\text{PDF} = 0.156(4)^{+(20)}_{-(6)}$,
with the error band in Fig.~\ref{fig:momenta-stacked} representing the sum in quadrature of the statistical and systematic uncertainties.

The phenomenological results thus obtained from the global QCD analyses lie between the DIS data extraction and the lattice QCD calculations.
This may suggest that DIS data are in mild tension with the DY and jet data from proton-proton collisions that have been included in the global analyses to complement the DIS data.
On the other hand, the tension between lattice simulations and phenomenological results may be milder than that indicated by comparison with the data extractions alone.

\section{Conclusions}
\label{sec.conclusions}

We have performed a detailed reanalysis of the world's inclusive DIS data, simultaneously obtained on protons and deuterons, to extract the structure function $F_2^n$ of the free neutron.
To account for the nuclear effects in the deuteron, we have consistently applied the nuclear correction calculation from the CJ15 global QCD analysis~\cite{Accardi:2016qay}, which includes calculated nuclear smearing and, in order to minimize the theoretical uncertainties, fitted nucleon off-shell corrections.
Special attention has been devoted to the normalization of the proton and deuteron experimental datasets and to the treatment of correlated systematic errors, as well as the quantification of procedural and theoretical uncertainties.
The data themselves have been carefully cross-normalized and shifted point-by-point, as allowed by the correlated systematic uncertainties, which turns out to be essential for minimizing the uncertainties in the extracted $F_2^n$.

As applications of the extracted neutron structure function, 
we considered a re-evaluation of the Gottfried sum rule, including for the first time its $Q^2$ dependence, and extracted the nonsinglet $F_2^p - F_2^n$ moment which provides an experimental benchmark for precision lattice QCD simulations.
In both cases, we have carefully evaluated the statistical and systematic uncertainties. 
The $Q^2$ dependence was found to be rather flat, within the uncertainties, indicating strong cancellations of higher-twist power corrections for the proton and neutron.

A comprehensive database including the world data on inclusive DIS from proton and deuteron targets, as well as the extracted neutron structure function and neutron-to-proton ratio, has been made publicly available (see Appendix~\ref{app:CJdatabase}).
To facilitate replication of our study and a comparison with other nuclear correction models, as well as for general applications, we also provide precomputed $(x,Q^2)$ grids in LHAPDF format for calculating the modified CJ15 PDFs used in our study, named \texttt{CJ15nlo\_mod}, and the corresponding DIS structure functions (see Appendix~\ref{app:F2grids}).
As a demonstration of the use of the DIS structure function grids, we illustrate neutron excess correction ratios for neutral and charged current DIS on nuclear targets (see Appendix~\ref{app:isoscalar}).
We expect these resources will be useful in future phenomenological applications requiring data on the neutron $F_2^n$ structure function.

\begin{acknowledgements}
We thank J.~Arrington, R.~Bhalerao, M.~E.~Christy, M.~Constantinou, H.-W.~Lin, J.~Morfin, E.~Segarra, and A.~Schmidt for useful discussions and communications.
This work was supported by U.S. Department of Energy contract DE-AC05-06OR23177, under which Jefferson Science Associates LLC manages and operates Jefferson Lab, DOE contracts DE-SC0008791, DE-AC02-05CH11231, and DE-FG02-88ER40410, and NSF grants number 1913257 and 2209995.
The work of WM was partially supported by the University of Adelaide and the Australian Research Council through the Centre of Excellence for Dark Matter Particle Physics (CE200100008).
\end{acknowledgements}

\begin{appendix}

\section{The CTEQ-JLab DIS database}
\label{app:CJdatabase}

Starting from the existing experimental datasets used in the CJ15 global QCD analysis~\cite{Accardi:2016qay}, a comprehensive database was constructed of unpolarized DIS measurements with proton and deuteron targets, along with the neutron target structure functions extracted in this analysis. 
The experimental observables included in the database are:
\begin{itemize}
    \item the $F_2$ structure functions, and their $d/p$ and $n/d$ ratios,
    \item the longitudinal to transverse cross section ratio $R=\sigma_L/\sigma_T$,
    \item the reduced cross section 
        $\sigma_{\rm red}(x,Q^2) = F_2(x,Q^2) - (y^2/Y^+)\, F_L(x,Q^2)$,
\end{itemize}
where $y=\nu/E$ is the lepton inelasticity, and $Y^+ = 1 + (1-y)^2 + 2M^2 x^2 y^2/Q^2$, with $E$ the incident lepton energy and $\nu$ the energy transfer in the target rest frame.
The reduced cross section datasets were extracted from experimental cross sections according to
\begin{eqnarray*}
\sigma_{\rm red}(x,Q^2) 
&=& \frac{x\, Q^4}{2\pi \alpha^2 Y^+} \frac{\dd^2\sigma}{\dd{x} \dd{Q^2}}.
\end{eqnarray*}
For the first time we have also included the full DIS datasets from the JLab~6~GeV program, which expanded the kinematic coverage in the high-$x$ region (see Figs.~\ref{figure:f2-p} and \ref{figure:f2-d} below).
A list of included experiments and observables is shown in Table~\ref{tab:DIS-data}.
The $F^n_2$ and $F^n_2/F^p_2$ data points were extracted, respectively, from the matched proton and deuteron data and from the measured $d/p$ and $n/p$ ratios (see Sec.~\ref{sec:pd_cross_normalization}). 
Extensive efforts were made to collect details of normalization, and correlated and uncorrelated systematic uncertainties on all the datasets.

\begin{figure}[htb]
  \centering
  \includegraphics[width=0.85\linewidth]{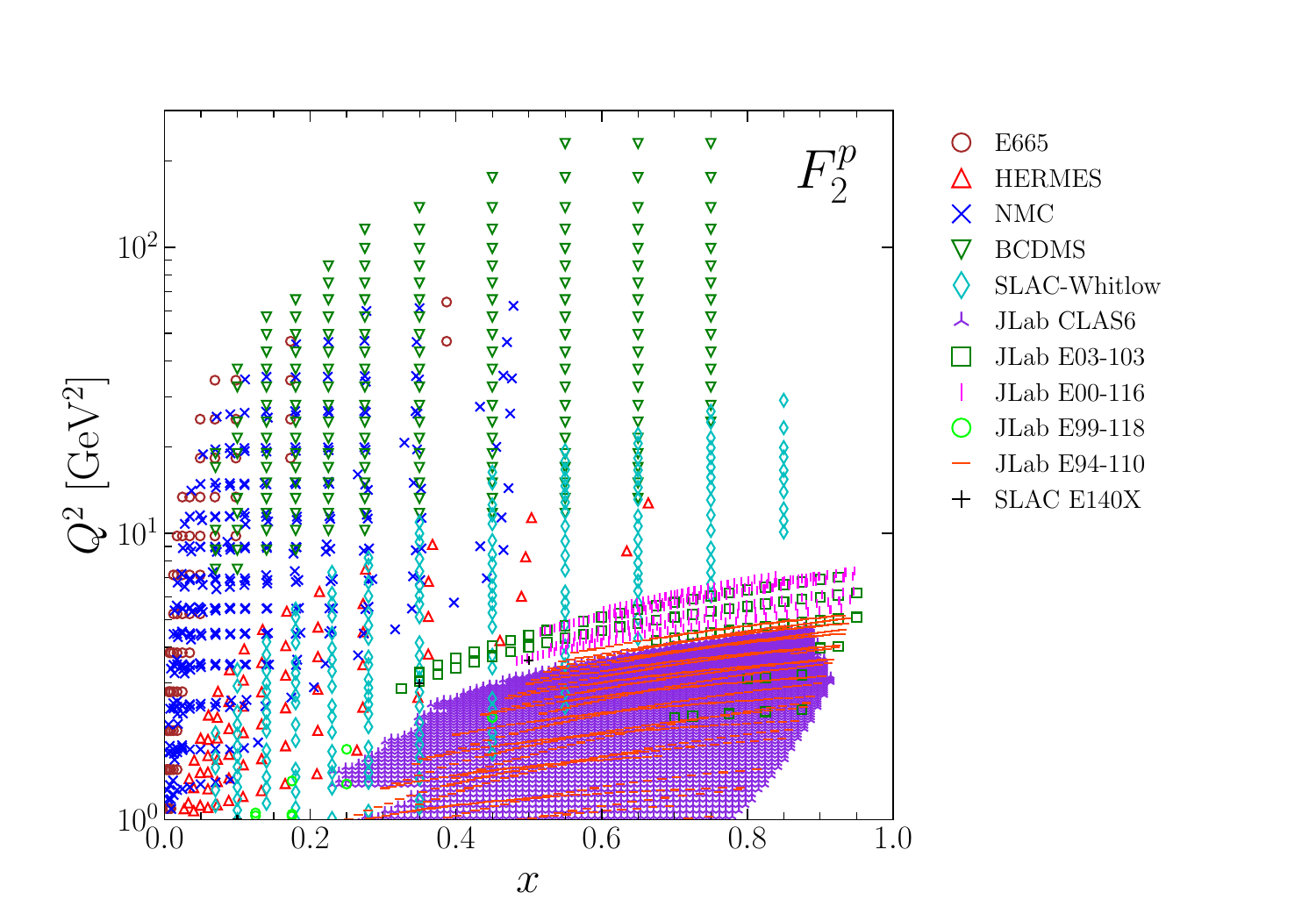}
  \includegraphics[width=0.85\linewidth]{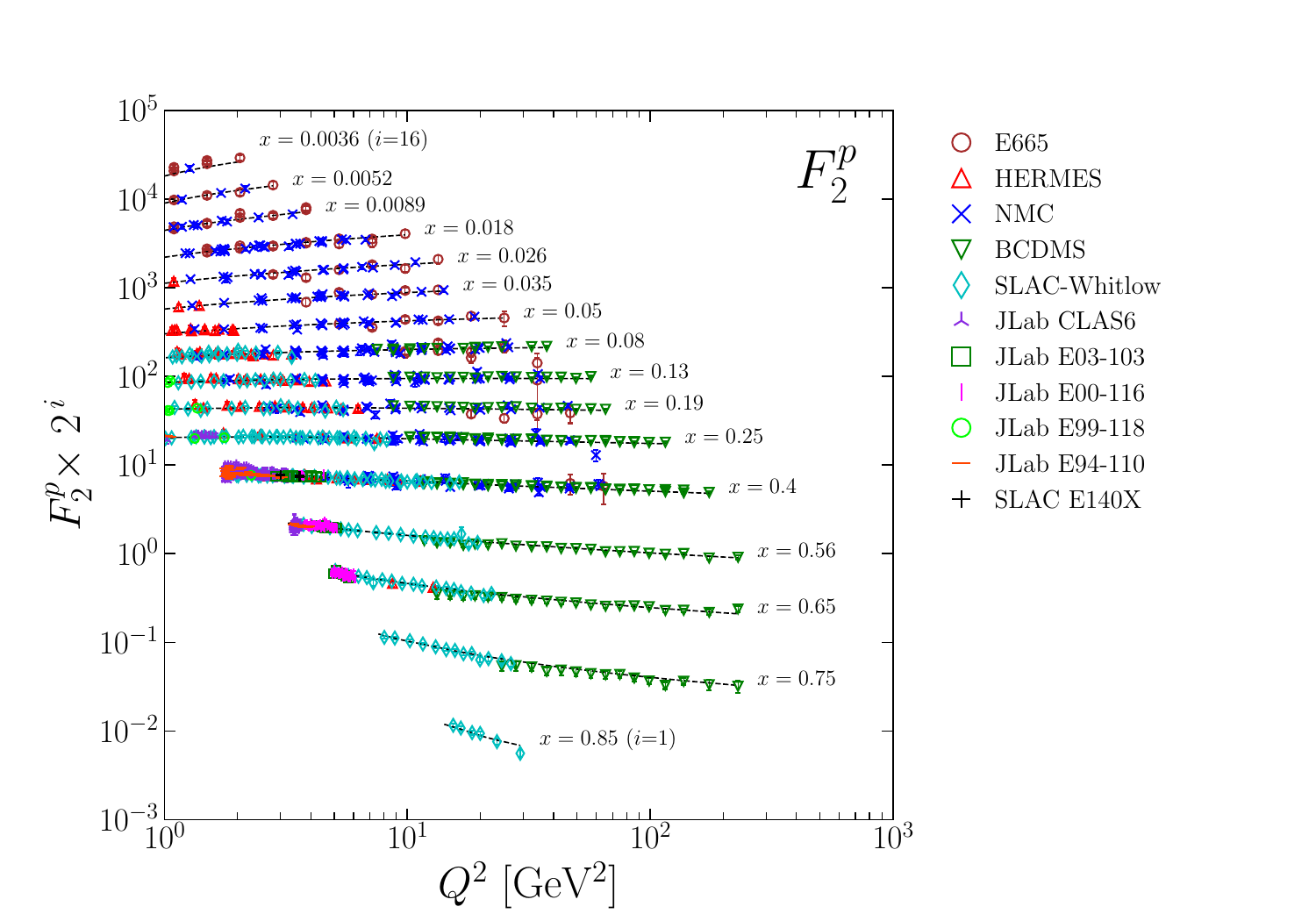}
  \caption{\setstretch{1.0} Kinematic range (upper panel) and proton $F_2^p$ data included in this database, binned in~$x$ (lower panel). The proton $F_2^p$ values are cut on $W^2=3.5$~GeV$^2$. Note that the available $\sigma_{\rm red}$ data will have a different (larger) kinematic coverage.}
  \label{figure:f2-p}
\end{figure}

\begin{figure}[htb]
  \centering
  \includegraphics[width=0.85\linewidth]{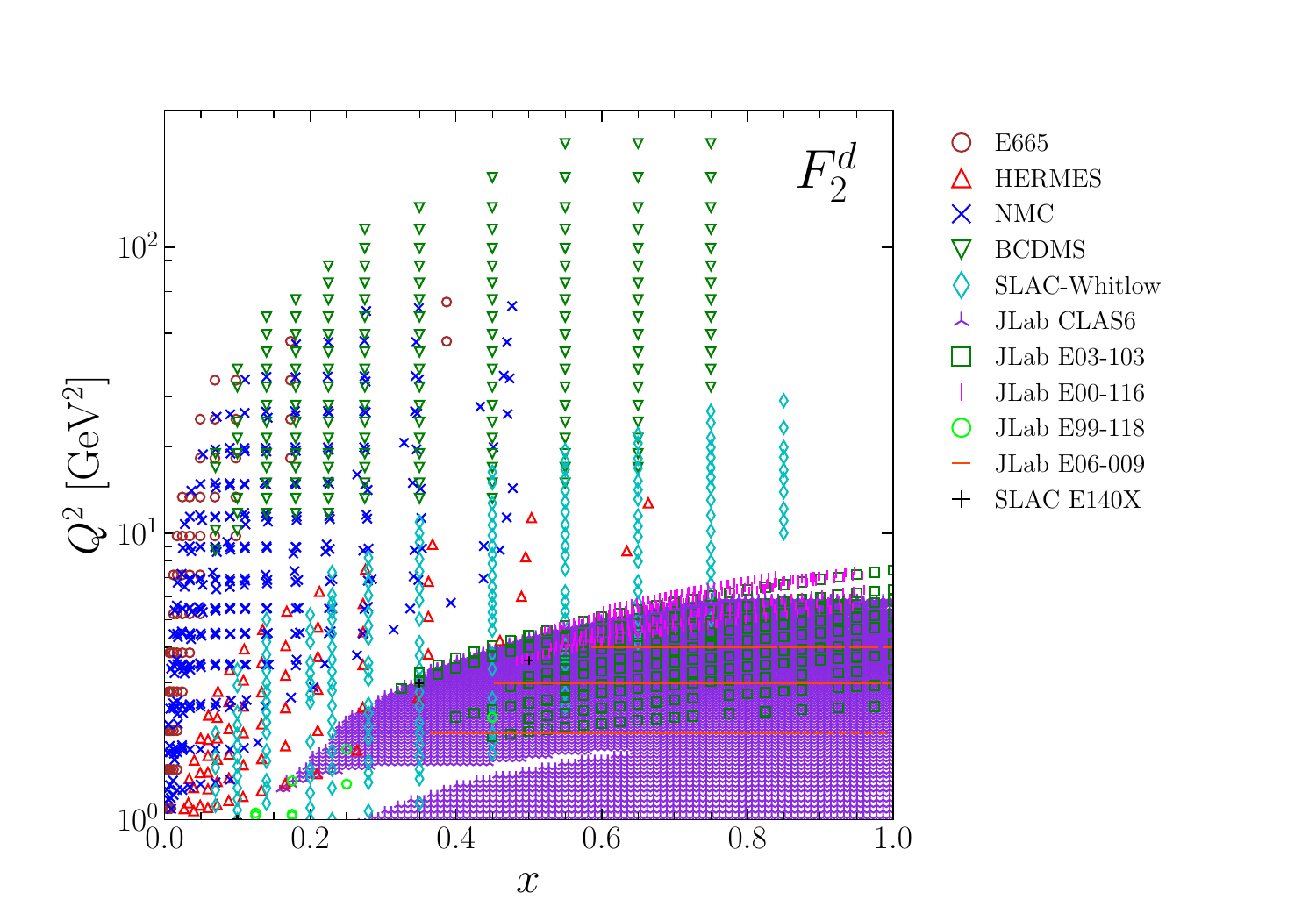}
  \includegraphics[width=0.85\linewidth]{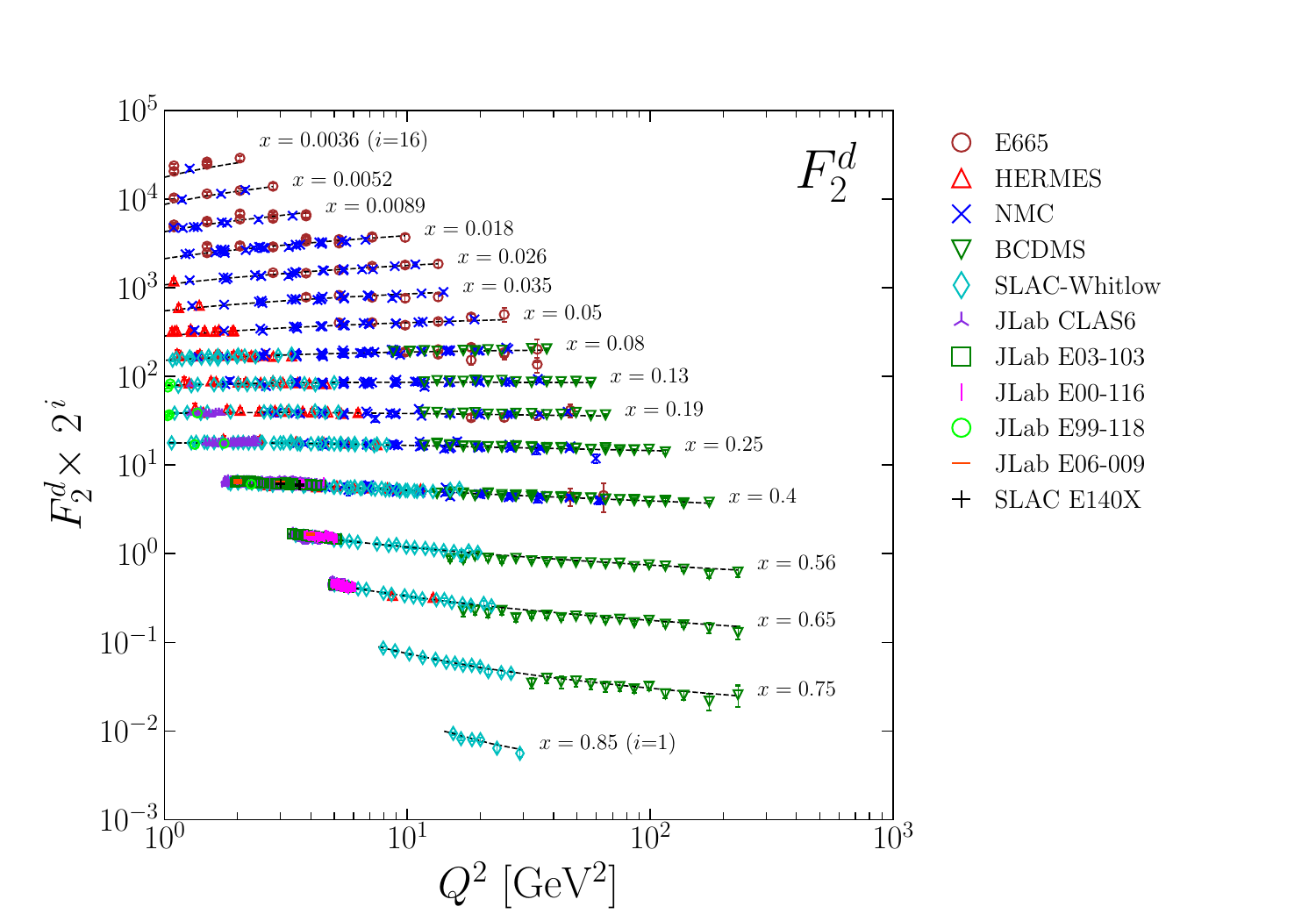}
  \caption{\setstretch{1.0}  
  As in Fig.~\ref{figure:f2-p}, but for the deuteron $F_2^d$ data.}	
  \label{figure:f2-d}
\end{figure}

All datasets are maintained in a public GitHub repository~\cite{CJ-git} in both Excel and plain text formats (the latter compatible with the CJ15 fitting package).
A 5-digit identifier is assigned to each dataset, and the content, references, source of uncertainties, and related calculations are documented in each README file.

\begin{table}[tbh]
\caption{\setstretch{1.0} List of experiments and observables included in the DIS database. Datasets marked with ``$*$'' were also included in the CJ15 global QCD analysis. The extracted $F_2^n$ and $R_{n/p}$ are also provided.\\}
\label{tab:DIS-data}
\begin{tabular}{l|c|c|c|c|c}
\hline
~Experiment   
& $\sigma_{\rm red}$ 
& $F_2$   
& $\sigma_L/\sigma_T$ 
& $F_2^n$ extracted  
& $R_{n/p}$ extracted 
\\ \hline
~SLAC-Whitlow \cite{Whitlow1990}
& ~$p,d,d/p$~   & ~$p^*,d^*$~   & $p,d$     & $\checkmark$  & $\checkmark$
\\
~SLAC-E140 \cite{e140r}    
&               &               & $d$       &               &                       
\\
~SLAC-E140x \cite{e140x}  
& $p,d$         & $p,d$         & $p,d$     & $\checkmark$  &                  
\\
~NMC \cite{Arneodo1997a, Arneodo1997}   
& $p,d,d/p^*$   & $p^*,d$       &           & $\checkmark$  & $\checkmark$    
\\
~BCDMS \cite{Benvenuti1989, Benvenuti1990} 
& $p,d$         & $p^*,d^*$     & $p,d$     & $\checkmark$  &             
\\
~JLab E06-009 \cite{e06009} 
& $d$           & $d$           &           &               &
\\
~JLab E94-110 \cite{e94110liang, e94110eric} 
& $p$           & $p$           &           &               &
\\
~JLab E03-103 \cite{e03103, 03103thesis}
& $p,d$         & $p,d$         &           & $\checkmark$  &
\\
~JLab E99-118 \cite{e99118_f2, Tvaskis2010, Tvaskis2007}~ 
& $p,d,d/p$     & $p,d,d/p$     &           & $\checkmark$  & 
\\
~JLab JLCEE96 \cite{Niculescu1999} 
& $p$           &               &           &               &          
\\
~JLab E00-116 \cite{e00116_f2, Malace2009} 
& $p,d$         & $p^*,d^*$     &           & $\checkmark$  &         
\\
~JLab CLAS6 \cite{Osipenko2003, clas_d, Osipenko2003a, Osipenko2005}  
& $p,d$         & $p,d$         &           &               &   
\\
~JLab \bonus \cite{CLAS:2014jvt, bonus_emc}  
&               & $n, n/d^*$    &           &               & $\checkmark$    
\\
~HERA I+II \cite{Abramowicz:2015mha}    
& $p^*$         &               &           &               &
\\
~HERMES \cite{hermes}  
& $p,d,d/p$    & $p^*,d^*$      &           & $\checkmark$  & $\checkmark$   
\\
~E665 \cite{E665}
&               & $p,d$         &           &               &             
\\
\hline
\end{tabular}
\end{table}

\newpage
\section{The \texttt{CJ15nlo\_mod} PDF and structure function grids}
\label{app:F2grids}

The careful evaluation of theoretical uncertainties in the extraction of the neutron structure functions in the main text necessitated the repeated evaluation of NLO $F_2$ structure functions using a modified version of the 49 member strong \texttt{CJ15nlo} PDF set, which represents uncertainties stemming from variations of 19 PDF parameters, 2 off-shell parameters, and 3 HT parameters \cite{Accardi:2016qay} with a nominal 90\% confidence level \cite{Accardi:2016qay}. As a result of this effort we present here the \texttt{CJ15nlo\_mod} set of PDFs, as well as a corresponding set of calculated DIS structure functions. 

The modification of the published CJ15 PDF error set was needed to take care of deviations from the assumed Gaussian behavior of PDFs and observables around the fitted parameter values in the least constrained regions of parameter space, such as for observables sensitive to the $d$ quark at large $x$ values. 
We have for example observed non-negligible deviations from the assumed quadratic behavior of the $\chi^2$ function in the parameter subspace spanned by the power correction parameters, that in turn is correlated with the $d$ quark parameters governing its large momentum behavior. As a result, the curvature of the $\chi^2$ function in the vicinity of the minimum is typically underestimated by the eigenvalues of the numerically evaluated Hessian matrix, which then does not faithfully capture the global fit uncertainties.
One consequence of relevance for the present paper, is that the uncertainties of the $R_{d/N}$ ratio calculated with the master formula \eqref{eq:sym_err_norm} and the published error PDF set, are typically smaller than those displayed in the figures of Ref.~\cite{Accardi:2016qay} that, for simplicity, used the Hessian approximation for the structure function ratios. 

In order to compensate for the observed deviations from the expected Gaussian behavior, we have then followed the procedure discussed in Refs.~\cite{Accardi:2021ysh,HuntSmith:2022ugn}. Namely, we have rescaled the \texttt{CJ15nlo} PDF error sets along each eigendirection to ensure that they produce a $\Delta\chi^2 = 1.646$ in the fitted datasets compared to the best fit CJ15nlo PDF set. The rescaling factors are close to 1 in the majority of the cases, except indeed for the least constrained eigendirections, but produce non-negligible effects, for example, for the evaluation of the uncertainties of the $R_{d/N}$ ratio. Further discussion of Gaussian deviations and possible remedies can be found in the aforementioned references.

Finally, to facilitate the reproduction of our results, as well as for general use, we have made publicly available on the CJ website \cite{CJ-web} the modified 49-member \texttt{CJ15nlo\_mod} PDF set, as well as the corresponding proton, neutron and deuteron \texttt{CJ15nlo\_mod\_SF} structure functions, as precomputed $x,Q^2$ grids on the CJ collaboration's web page. 
All grids are provided in LHAPDF format~\cite{LHAPDF6} for easy Python and C access, and will soon be submitted for inclusion in the official LHAPDF website~\cite{LHAPDF-web}. The $F_2$ neutral current structure functions are presented both with and without TMC and HT corrections. The $F_L$ neutral current structure function and the charged current structure functions are only available at leading twist, since the CJ15 fit lacked the data to constrain their power corrections.
The indexing of structure functions within each LHAPDF grid follows the conventions laid out for the inclusive DIS studies presented in Ref.~\cite{AbdulKhalek:2021gbh} and is explicitly discussed in Ref.~\cite{txgrids}.

\section{Isoscalar corrections with \texttt{CJ15\_mod} structure functions} 
\label{app:isoscalar}

When comparing the experimentally measured lepton-nucleus cross sections for different nuclear targets, 
a correction to account for the proton and neutron number imbalance (or target non-isoscalarity) is often necessary to isolate dynamical nuclear effects that go beyond the trivial difference between the proton and neutron scattering cross section contributions. 
This process has been widely discussed in the literature, and we briefly rederive it here before discussing the use of the \texttt{CJ15nlo\_mod} proton and neutron structure functions, described in Appendix~\ref{app:F2grids}, for calculating the needed correction factor.

For a nucleus $A$ with $Z$ protons and $N$ neutrons, the nuclear effects in the lepton-nucleus cross section can be quantified by considering the ratio 
\begin{align}
    R_A^\text{nuc} = \frac{\sigma_A}{Z\sigma_p+N\sigma_n}
\label{eq:nucl_ratio}
\end{align}
of the nuclear cross section $\sigma_A$ to the sum of $Z$ free proton and $N$ free neutron cross sections, denoted by $\sigma_p$ and $\sigma_n$, respectively.
(The subscript $A$ label the nucleus under consideration, rather than the specific atomic number.)
It is then straightforward to obtain
\begin{align}
    R_A^\text{nuc} 
    = \frac{\sigma_A\, f_A^\text{iso}}{\frac12 A (\sigma_p + \sigma_n)},
\label{eq:nuclear_effect_ratio}
\end{align}
where the so-called ``isospin correction factor,''
\begin{eqnarray}
    f_A^\text{iso} 
    &=& \Big( \frac{A}{2} \Big)
    \frac{1 + \sigma_n/\sigma_p}{Z + N\sigma_n/\sigma_p} \ ,
\label{eq:fiso}
\end{eqnarray}
only depends on the ratio of free neutron to free proton cross sections.
Note that the denominator of the right hand side of Eq.~\eqref{eq:nuclear_effect_ratio} is just a scaled, isospin-symmetric free nucleon target cross section.
By analogy, one can then interpret
\begin{align}
    \sigma_A^\text{iso} \equiv \sigma_A\, f_A^\text{iso}
\end{align}
as an isospin-symmetrized nuclear cross section, with $f_A^\text{iso}$ providing the conversion from the measured $\sigma_A$ to the isoscalar cross section, $\sigma_A^\text{iso}$.
Having thus removed the trivial nuclear effects due to the number imbalance of protons and neutrons in a given nucleus, one can compare and contrast nuclear effects in different nuclei on the same footing, such as in the double ratios $(\frac{1}{A}\sigma^\text{iso}_A)/(\frac{1}{B}\sigma_B^\text{iso})$ of the symmetrized cross sections for nuclei $A$ and $B$.

In practice, the cross section ratio data are often converted to structure function ratios, ${\sigma_A}/{\sigma_d} \approx {F_2^A}/{F_2^d}$, before applying the isoscalarity corrections, under the assumption that the longitudinal to transverse cross section ratio \mbox{$\sigma_A^L/\sigma_A^T$} is independent of the nuclear target. 
Repeating the derivation above, one can see that the isoscalar correction factor $f_A^\text{iso}$ would analogously apply to the nuclear structure function. 
However, one should keep in mind that, while a relatively small effect, the $A$ independence assumption for the longitudinal to transverse ratio may not be accurate at large $x$ and low $Q^2$ values~\cite{Guzey:2012yk, Tvaskis2010, Niculescu1999}.

Experimentally, in the absence of free neutron targets, the nuclear modification ratio \eqref{eq:nucl_ratio} is often approximated by the ``nuclear EMC ratio,"
\begin{align}
    R_A^\text{EMC} = \Big( \frac{2}{A} \Big) \frac{\sigma_A^\text{iso}}{\sigma_d},
\end{align}
of the per-nucleon nuclear and deuteron cross sections, with the isoscalar correction applied to the numerator (although sometimes this correction is not applied).
A nonzero nuclear modification in the deuteron, $R_d^\text{nuc} \not= 1$, was recently observed by the BONuS Collaboration~\cite{bonus_emc}, with further precision measurements planned~\cite{bonus12}, and shown to be non-negligible especially at large values of $x$ (see also Fig.~\ref{fig:dN-CJ-AKP-BoNUS}).
The neutron data extracted in Ref.~\cite{bonus_emc} obviate the need for using deuteron targets as proxies for the combination of free proton and neutron targets, enabling one to directly measure $R_A^\text{nuc}$ and fully expose the nuclear effects in the nuclear target.

\begin{figure}[tb]
  \centering
  \includegraphics[width=0.47\linewidth]{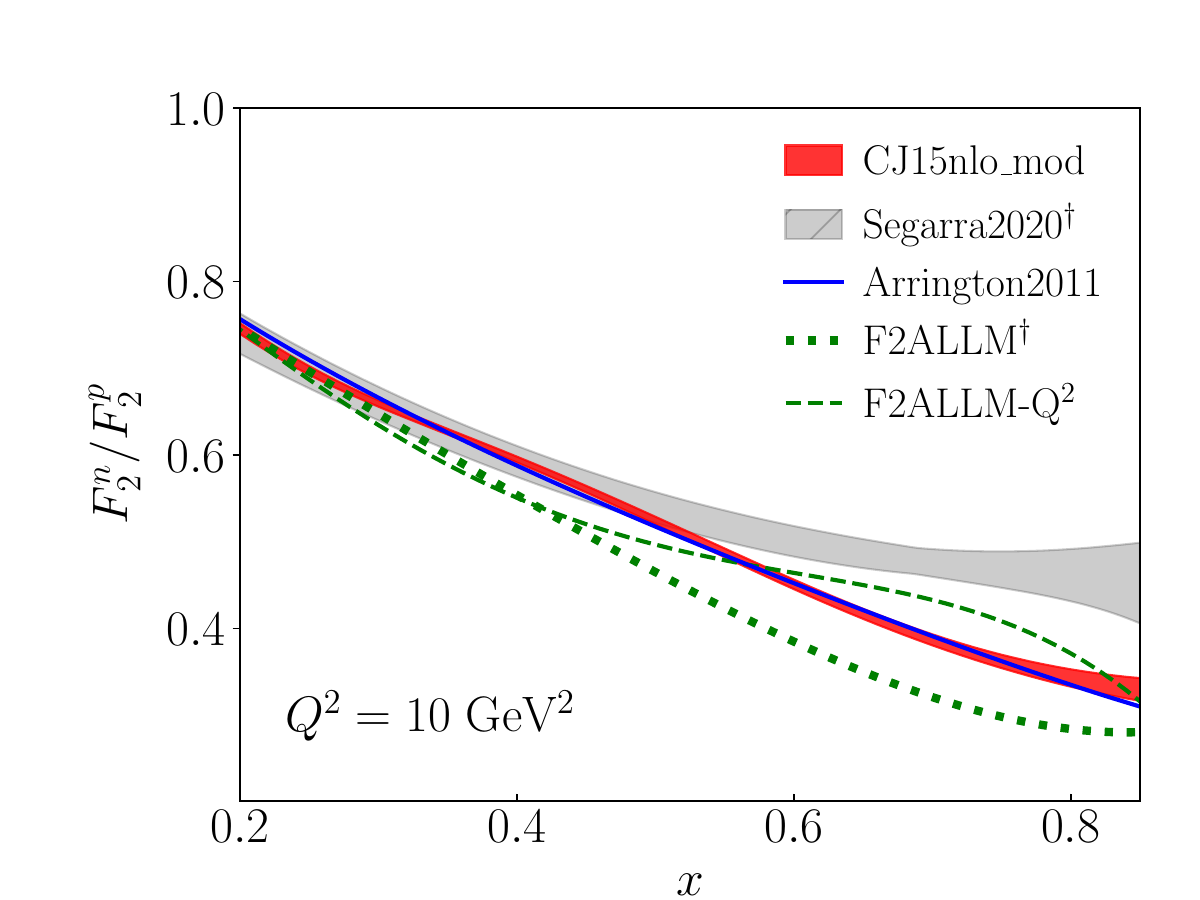}\hspace*{-0.2cm}
  \includegraphics[width=0.47\linewidth]{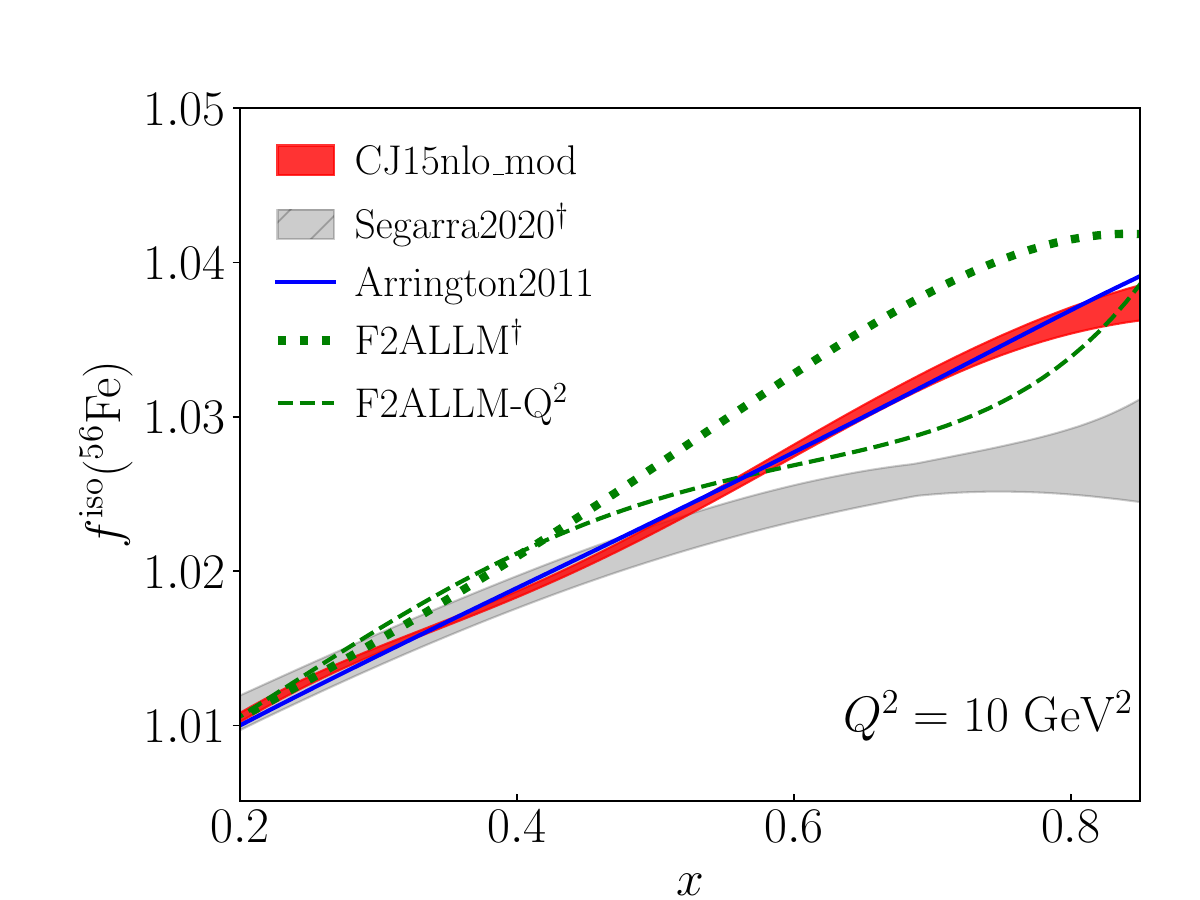}
	  \caption{\setstretch{1.0} Neutron to proton $F_2$ structure function ratio (left) and isoscalar corrections for $^{56}$Fe (right) at $Q^2=10$~GeV$^2$ from this work and several empirical fits including the F2ALLM model with and without $Q^2$ dependence~\cite{nmc_f2allm}, and more recent fits by Arrington~{\it et~al.}~\cite{Arrington:2011qt}, and Segarra {\it et~al.}~\cite{Segarra:2019gbp}. Models with $Q^2$ dependence are labeled by $\dagger$.}
  \label{figure:np-ratio}
\end{figure}

The \texttt{CJ15nlo\_mod} proton and neutron structure functions obtained in this work (see Appendix~\ref{app:F2grids}) can be directly used to calculate the isoscalar correction factor $f_A^{\rm iso}$ at any $x$ and $Q^2$.
Indeed, for events with not too large lepton energy loss, one can approximate
\begin{equation}
    {\frac{\sigma_n}{\sigma_p}} \approx {\frac{F_2^n}{F_2^p}},
\end{equation}
and obtain
\begin{equation}
    f_A^{\rm iso}
    \approx \Big( \frac{A}{2} \Big)
    \frac{1 + F_2^n/F_2^p}{Z + N F_2^n/F_2^p},
\label{eq:fisoF2}
\end{equation}
which depends only on the $F_2^n/F_2^p$ ratio. 
(Note that the CJ15 PDF fit from which the \texttt{CJ15nlo\_mod} is derived assumes isoscalar, multiplicative higher twist corrections. For determination of $f_A^{\rm iso}$ one can therefore use the leading-twist structure functions that were perturbatively calculated at NLO using the  \texttt{CJ15nlo\_mod} PDF sets. This is important when dealing with charged current DIS data, which were not included in the CJ15 analysis.)

In Fig.~\ref{figure:np-ratio}, the \texttt{CJ15\_mod} $F_2^n/F_2^p$ ratio is compared to several widely used empirical fits. 
\texttt{F2ALLM}~\cite{nmc_f2allm} fitted $F_2$ data prior to 1997 above the resonance region, and is provided with and without $Q^2$ dependence.
\texttt{Arrington2011}~\cite{Arrington:2011qt} used a similar deuteron smearing and off-shell corrections as CJ15 and results in a similar $F_2^n/F_2^p$ ratio. 
\texttt{Segarra2020}~\cite{Segarra:2019gbp} assumed a universal modification of nucleons in short-range correlated (mostly neutron-proton) pairs, and included recent light nuclear target data, as well as data from heavy nuclei, in their fits.  
In general, the $F_2^n/F_2^p$ ratio from all these models are in good agreement at small $x$, but show large systematic deviations at large $x$ stemming from the adopted neutron extraction procedure. 
Among the considered fits and parametrizations, \texttt{CJ15nlo\_mod} is the only $Q^2$-dependent model that also provides a rigorous uncertainty estimation.
The isoscalar correction factor for $^{56}$Fe in Fig.~\ref{figure:np-ratio} is calculated using Eq.~\eqref{eq:fisoF2} with different $F_2^n/F_2^p$ ratio models, and reveals a model dependence of the order of 10\% at large values of $x$.

\begin{figure}[tb]
  \centering
  \includegraphics[width=0.47\linewidth]{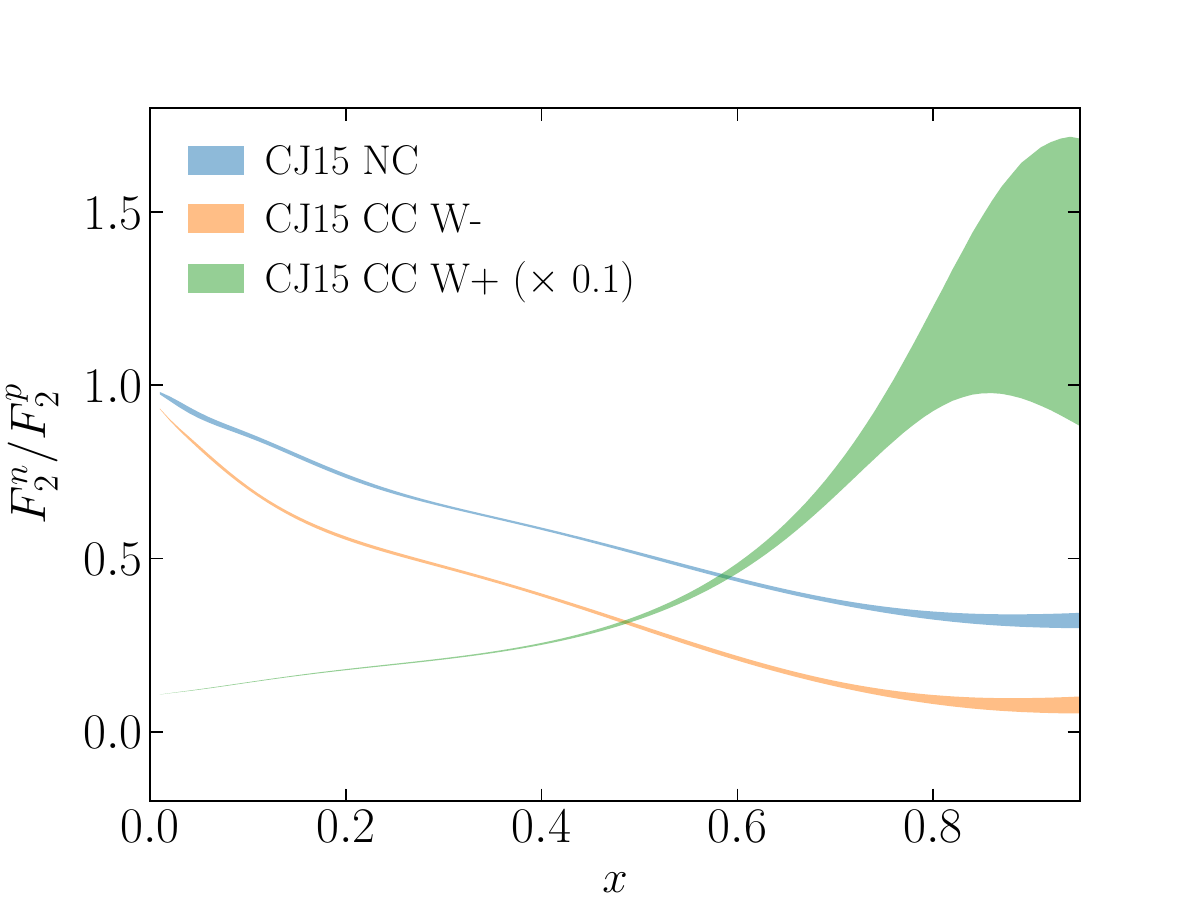}
  \includegraphics[width=0.47\linewidth]{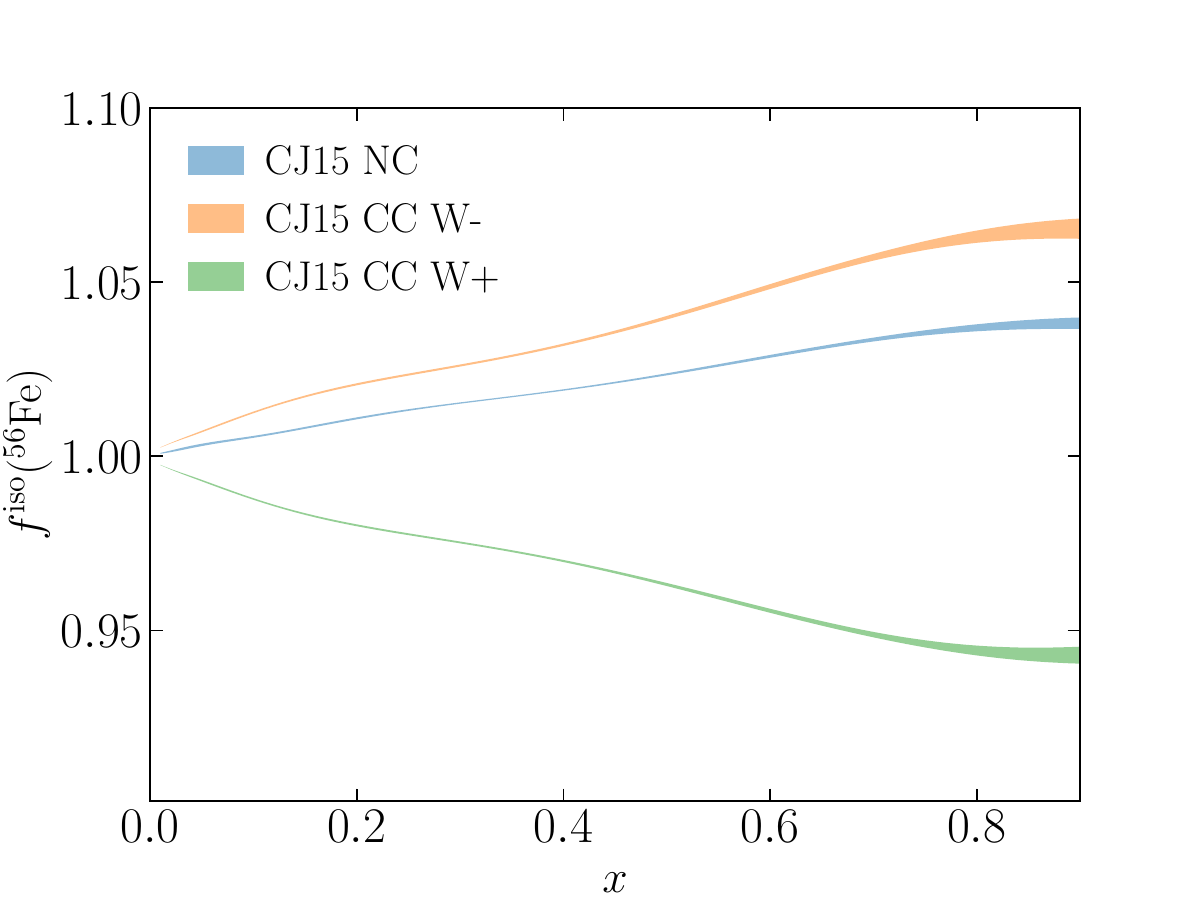}
  \caption{\setstretch{1.0} 
    (Left) Neutron to proton $F_2$ ratio calculated at $Q^2=20$ GeV$^2$ using CJ15 PDFs.  The positively charged CC $W^+$ ratio, that at LO is equal to the reciprocal of the negatively charged current (CC) $W^-$ ratio and grows as $x\to1$, is compared to the neutral current (NC) ratio and the CC $W^-$ ratio on a linear scale. 
    (Right) The isoscalar correction factor for $^{56}$Fe calculated from corresponding neutron to proton ratios. 
    }
  \label{fig:np-ratio-CC}
\end{figure}

The neutron to proton ratio may also be utilized in accounting for neutron excess effects in neutrino-nucleus scattering.
In contrast to electron scattering, here the neutrino-neutron cross section is actually larger than the neutrino-proton cross section.
The charged current (CC) neutron to proton ratio $R^{\rm CC}_{n/p}=F_2^{n,\rm CC} / F_2^{p,\rm CC}$, calculated using CJ15 PDFs, is shown in Fig.~\ref{fig:np-ratio-CC} for $W^-$ exchange ($e^-p \to \nu X$ and $\bar{\nu} p \to e^+ X$) and $W^+$ exchange ($e^+p \to \bar\nu X$ and $\nu p \to e^- X$). 
At leading order, those two ratios are related using isospin symmetry by
\begin{equation}
    R_{n/p}^{W^+} \approx \frac{1}{R_{n/p}^{W^-}}.
\end{equation}
In the $x \to 1$ limit, one finds
\begin{equation}
    R_{n/p}^{W^-} \to \frac{d}{u},
\end{equation}
which can be compared with the corresponding neutral current ratio,
        $R_{n/p}^{\rm NC} \to \frac14 (1 + \frac{15}{4} d/u)$.
The isoscalar corrections calculated with $F_2^{\rm CC}$ are of the same order as those from the NC case, as shown on the right panel of Fig.~\ref{fig:np-ratio-CC}. 

At large $x$ the $W^-$ ratio $R_{n/p}^{W^-}$ is essentially a shifted down version of the NC ratio, and decreases towards zero following the behavior of the fitted $d/u$ ratio.
Conversely, the ratio $R_{n/p}^{W^+}$ for $W^+$ exchange grows rapidly as $x \to 1$; this growth is tamed if the $d/u$ ratio tends to a finite limit, as in the CJ15 fit.
The $R_{n/p}^{W^+}$ structure function ratio therefore seems to be a particularly sensitive probe of the large-$x$ behavior of the $d$ and $u$ quark PDFs and of the nucleon's nonperturbative structure.
It could be measured in $e^+ + p(d) \to \bar\nu + X$ reactions with a positron beam at JLab12 \cite{Accardi:2020swt} and at the Electron-Ion Collider \cite{eic-yellow}, or in $\nu + p(d) \to e^- + X$ processes at high-energy neutrino facilities \cite{Petti:2022bzt}.

\end{appendix}

\bibliographystyle{apsrev4-1}
\bibliography{ref}
\end{document}